\documentclass[%
 reprint,
 showkeys,
 amsmath,amssymb,
 aps,
]{revtex4-1}
\usepackage{placeins}
\usepackage{graphicx}
\usepackage{dcolumn}
\usepackage{bm}
\graphicspath{{Images/}}

\begin{document}


\title{(Private Copy) Physics of Reverse Flow on Rotors at High Advance Ratios}
\thanks{This draft is a private copy}%

\author{Nandeesh Hiremath}
\author{Dhwanil Shukla}%
\author{Narayanan Komerath}
\affiliation{%
 Georgia Institute of Technology, School of Aerospace Engineering, Atlanta, Georgia, 30332-015, USA\\
}%

\date{\today}

\begin{abstract}
The reverse flow phenomenon on a retreating rotor blade at high advance ratios shows vortical structure. A differential onset velocity gradient due to the rotor rotation creates an envelope of reverse flow region. A Sharp Edge Vortex (SEV) is observed at the geometric trailing edge for such an edgewise flow. The SEV grows radially inwards in its size and strength as the pressure gradient inside the vortex core counteracts the centrifugal stresses. The attached SEV convects along with the rotor blade going through the phases of evolution and dissipation of the vortex. At an azimuth angle of 240 degrees, a coherent vortex begins to form at the start of reverse flow envelope extending radially inwards. The vortex size and strength increases as the blade traverses to 270 degree azimuth, followed by eventual cessation. The reverse flow envelope increases at higher advance ratios with an increased vortical strength. The solid core body of the vortex stretches at lower advance ratios relating to lower rotational energy, and shows the signs of a burst vortex in the dissipation phase. 

The proximity of the SEV to the rotor blade would create large excursions in the surface static pressures which in turn generates significant negative lift on the retreating rotor blade. The attached, coherent sharp edge vortex shows similar morphological features as the leading edge vortex on a delta wing. And the reverse flow region is hypothesized as analogous to the vortex lift generation on delta wings. 


\end{abstract}

\maketitle


\section{Introduction}
The edgewise flow on the retreating side of a rotor blade in high advance ratio ($\mu$) forward flight poses several aerodynamic phenomena. Radial flow, differential on-set velocity, centrifugal stresses, dynamic stall and reverse flow are some of the prominent features observed in this flow field. As the rotor moves into the retreating phase, in order to balance the rolling moments between the advancing and retreating sides, the blade pitch is increased. This results in substantial portions of the blade, especially the outboard regions experiencing stall, resulting in increased vibrations. Where the pitch rate is high, the blade retains dynamic lift well past the static stall angle of attack, and then experiences dynamic stall, with sharp excursions in pitching moment. The advancing flow on the outboard regions encounters very high pitch angles resulting in stall, and this effect increases in the inboard regions (outside reverse flow regime) due to differential on-set velocity. The far inboard regions of the blade experiences reverse flow as the rotation causes the obvious variation of the onset velocity from root to tip. With the increase in advance ratio, the reverse flow envelope increases. The reverse flow region experiences a strong inward radial flow which is dictated by a pressure gradient caused by three-dimensional vortex that counteracts the centrifugal effects at the rotor surface. 

\begin{figure}[!t]
\centering
\includegraphics[width =0.8\columnwidth]{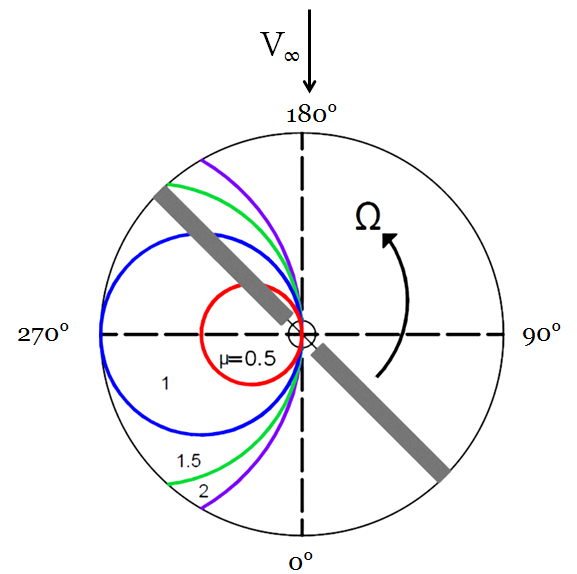}
\caption{Extent of reverse flow region with increasing advance ratio ($\mu)$}
\label{reverseregion}
\end{figure}

As the ratio of flight speed to rotor tip speed increases (increasing advance ratio), the retreating rotor encounters reverse flow on large portions of the rotor disk as shown in Figure \ref{reverseregion}. This phenomenon can also be observed on slowed rotors, such as on an autogyro or compound helicopter, or wind-turbine blades encountering edgewise gusts. This phenomenon leads to many uncertainties, one of which is observed in the location of airfoil center of pressure. For instance, in the inboard regions that first experience reverse flow, the center of pressure suddenly shifts to 75\% from 25\% of the chord as measured from the blunt edge, where as the outboard portion still experiences it at the quarter chord. In addition, the lift direction switches from up to down in the reversed flow regime. These induce both quasi-steady changes in bending and twisting moments on the blade, followed by vibratory loads caused due to the cyclic nature of the blade aerodynamic loading. These cyclically varying loads may lead to fatigue and eventual failure of the rotor blade and pitch links. Also, the blunt edge might cause high fluctuations in drag. These phenomena are becoming increasingly common in the reverse flow region with the advancement of the rotorcraft industry in high speed compound helicopters and co-axial counter rotating helicopters.

Prior approaches to analyzing reverse flow were from a two-dimensional airfoil / high aspect ratio wing viewpoint, applying yaw and aspect ratio corrections to 2D airfoil data. Such models predict that the region below the blade features a dead 2-D recirculation zone, and the blunt base has massive separation with vortex shedding. Even when sweep corrections are applied, history effects (birth and evolution of vortex) are not captured. From our published work \cite{44ShuklaASME2015, 2HiremathAHS-2015}, we argued that the formation and evolution of a strong three-dimensional (helical) vortex early on the retreating blade side, would be key to the entire problem. At azimuth angles between 181 degrees and 240 degrees, the sharp edge of the blade resembles the edge of a highly forward-swept wing, causing a strongly helical vortex to form, with significant axial flow in its core. As azimuth angle increases, the yaw angle decreases, and the subsequent evolution of the vortex bears much similarity to the observed phenomena on delta wings. We explored this hypothesis, that the genesis of the aerodynamic loads was best viewed through the perspective of vortex flows occurring on sharp-edged swept wings at an angle of attack. A rectangular untwisted NACA0013 rotor blade of (semi-span) aspect ratio 3.47 was tested as a fixed wing in a wind tunnel, operated through a large variation in yaw and smaller variations in angle of attack, in both the forward and reverse-facing orientations. With the root cutout included during operation as a rotor blade, this corresponds to an aspect ratio greater than 8 in airplane wing terms. The phenomena in reverse flow are dominated by the sharp-edge vortex which is generated along the sharp edge of the blade. This vortex is three-dimensional with high values of core axial velocity directed towards the inboard sections. It originates at the tip in the case of the forward quadrant, and at the root in the aft quadrant of the rotor. This flow field does not conform to 2-D airfoil aerodynamics, nor to the high-aspect-ratio corrections derived from Prandlt’s lifting line theory approximations, even for a full-scale rotor blade. The experiment here aims to reveal the detailed flow phenomena, as opposed to producing scale-model results. The rotor radius of 0.889 m was limited by the 2.74 m width and 2.13 m height of the wind tunnel where the rotor is operated. Any higher values of rotor radius would have created higher wall interference of tip vortex and wake. As a consequence of relatively low aspect ratio, a larger chord enables the Reynolds number based on chord to go well above the laminar regime. This is essential to avoid the laminar to turbulent transition on the surface. It also enables better resolution in capturing the velocity and eventually the pressure distributions on the blade.

\begin{figure}[!t]
\center
\includegraphics[width =1\columnwidth]{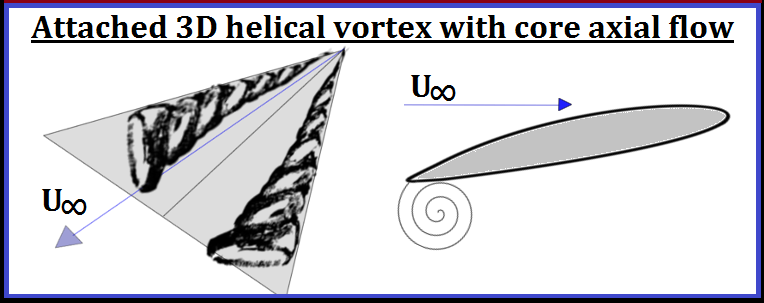}
\caption{Reverse flow hypothesis as a Delta Wing concept}
\label{CoverPage_Final}
\end{figure}

\section{Prior work}
The advance ratios experienced by the rotor blades of modern helicopters such as the Sikorsky X2 and Eurocopter X3 can exceed 0.8 \cite{5WalshAHS-2011, 6DattaJAHS-2013}.  At these high advance ratios, significant sections of the retreating blade experience reverse flow, characterized by the freestream hitting the geometric trailing edge of the blade first and traveling toward the geometric leading edge. The resultant flow field is not well understood and is characterized by early flow separation, negative lift, and periodic vortex shedding \cite{7LindAIAA-2013}. Reverse flow is a major limiter in the design of high-speed rotorcraft. Helicopter designs that attempt to mitigate the effects of reverse flow date back to the early 1970’s. Fairchild Republic Division’s Reverse Velocity Rotor \cite{8EwansDTIC-1973} concept featured a negative pitch angle on the retreating blade in order to achieve a positive angle of attack in the reverse flow regime, and incorporated airfoils with blunt leading and trailing edges. The more recent Sikorsky X2 Technology Demonstrator \cite{9BagaiAHS-2008} has coaxial rotors based on the Advancing Blade Concept \cite{10RuddellDTIC-1981} platform and also features blunt trailing edge airfoil sections near the blade root. 

As the rotor's advance ratio increases, the region of reverse flow on the rotor disk expands, reaching the blade tip at $270^\circ$ azimuth at an advance ratio $\mu$ $\approx$ 1. Prior work has been surveyed in \cite{2HiremathAHS-2015} and is summarized here. The most basic method was to take airfoil data \cite{7LindAIAA-2013,11Knight-1929,12CritzosNACA1955,13JacobsNACA-1937,14Bradley-1956} from 2D wind tunnel tests in the region around 180 degrees angle of attack, and correct for yaw with a sinusoidal correction, and if needed, an aspect ratio correction. At the next level, with experiments revealing evidence of vortices in the flow field, `vortex shedding’ was postulated; both from the blunt edge when using NACA0012 type airfoils with blunt leading edges, and from the sharp edge as from a thin flat plate held at negative angle of attack \cite{7LindAIAA-2013}. Karman vortex street analogues were also developed. These models have recently been adapted to include unsteady pitching effects, again from airfoil experimental data. Other studies with unswept airfoils have modeled the sharp-edge flow as a diffused, separated shear layer, with the blunt edge shedding vortices. Others have postulated a `reverse-chord dynamic stall’ process \cite{6DattaJAHS-2013}, explaining the occurrence of sharp pitching moments as well as strong vortices seen in the flow field in experiments. 

The experimental work performed by Wheatley \textit{et al}. \cite{15WheatleyNACA1936} showed force measurements on a Pitcairn PCA-2 autogyro rotor at various pitch settings and advance ratios up to $\mu$= 0.7 and found a negative correlation between lift coefficient and advance ratio. Charles \textit{et al}. \cite{16CharlesDTIC-1969} tested a UH-1D rotor at advance ratios up to $\mu$ = 1.1 and found that rotor performance predictions broke down at $\mu >$ 0.5. They also observed flapping instability and a long transient response to control input at $\mu$ = 1.1. MacCloud \textit{et al}. \cite{17MacCloudNASA-1968} performed tests on a teetering rotor with a NACA 0012 airfoil at advance ratios up to 1.0 and observed a drop in lift coefficient at high advance ratios. Harris \textit{et al}. \cite{18HarrisNASA-2008} performed correlation studies on these early datasets in 2008 and concluded that 3D Navier-Stokes solver OVERFLOW-2 could not accurately predict lift, drag and pitching moments for reverse flow over airfoils and could not be recommended for use at an advance ratio beyond 0.35. Other recent work includes \cite{18HarrisNASA-2008,19QuackenbushAHS-2011,20PostdamAHS-2012,21DattaJAC-2006, 22FlaxJAC-2012,23NiemiecSMS-2014,24Carter-2001}. The survey by Quackenbush \cite{19QuackenbushAHS-2011}, though designed for computations, captures many of the phenomena to be expected from the reverse flow geometry, and even shows a sharp-edged vortex. Harris \cite{18HarrisNASA-2008} summarizes high advance ratio rotorcraft flight experience. Niemiec \cite{23NiemiecSMS-2014} and Carter \cite{24Carter-2001} describe airfoils intended for the reversed-flow regime. Uncertainties in lift, drag, pitching moment, blade bending and twist and rotor stability are cited in \cite{25MeyerDTIC-1953, 26SweetNACA-1964}. Slowed-rotor compound rotorcraft with other lifting surfaces to offload the rotor in high-speed flight are reported in \cite{8EwansDTIC-1973, 15WheatleyNACA1936, 16CharlesDTIC-1969, 17MacCloudNASA-1968, 27Hislop-1958, 28WheatleyNACA-1936, 29LandgrebeAHS-1977, 30FlorosAHA-2009, 31Newman-AEAT1997, 32WarwickFI-2005, 33HarrisAHS-1970}, and more modern designs in \cite{5WalshAHS-2011, 34BagaiAHS-2008}. 

A recent study on a slowed UH-60A rotor performed by Datta \textit{et al}. \cite{6DattaJAHS-2013, potsdam2016computational, norman2011full, floros2009performance, berry2012performance} showed evidence of reverse chord dynamic stall and large pitch-link load impulses on the retreating side of the rotor disk at advance ratios near unity. Kottapali \cite{35Kottapalli-2012, kottapalli2011enhanced} made initial attempts to predict the blade loads from this study using CAMRAD II, but they showed serious issues at $\mu > 0.8 $. Predictions made by Yeo \cite{36YeoJA-2012} using CAMRAD II showed fair airload and structural load correlation. Potsdam \textit{et al}. \cite{20PostdamAHS-2012, ortun2016rotor} performed coupled CFD and comprehensive analysis on this dataset and predicted unconventional wake patterns and a lower surface vortex on the retreating blade, attributing that to dynamic stall. Pitching moment predictions on the advancing and retreating sides were not encouraging. Lee \textit{et al} \cite{park1994numerical} performed time-averaged force measurements and flow visualization on various airfoil sections in reverse flow and found a drag jump at $\alpha$ =$180^\circ$ due to the unsteady formation and convection of a large vortex in the wake. Lind \textit{et al}. \cite{7LindAIAA-2013, lind2015experimental, lind2016unsteady, lind2016reynolds, lind2015reynolds} also conducted reverse flow studies on static sharp and blunt trailing edge airfoils. They found that the large, negative angles of attack of the inboard section of a rotor in reverse flow causes flow separation and the onset of vortex shedding. It was suggested that the use of a blunt-trailing edge airfoil with a relatively linear lift curve slope would be ideal for the reverse flow regime. However, Lind \textit{et al}. did not study the reverse flow aerodynamics of static yawed airfoils or finite wings. Ormiston et.al. \cite{ormiston2008new, ormiston2012rotor, ormiston2009analytical} performed a computational analysis on the rotor blades at high advance ratios, and showed inconsistency with the observed experimental data due to limited knowledge of the aerodynamic model. 

In summary, the past approaches to the rotor reverse flow problem have focused on the 2-D airfoil model. The 270-degree azimuth angle has usually been taken as the starting point for analyses, where the blade has no yawed flow. To explain the vortical structures reported by experimenters, computational researchers have focused on pitching and plunging airfoils, even venturing into `reverse chord dynamic stall’ \cite{berry2014slowed,berry2011wind,berry2013high,bowen2014validation}. The presumption appears to have been that the rest of the retreating blade sector can be analyzed as perturbations of the 270-degree case, with yaw corrections applied.

\begin{figure}[!t]
	\center
	\includegraphics[width =1\columnwidth]{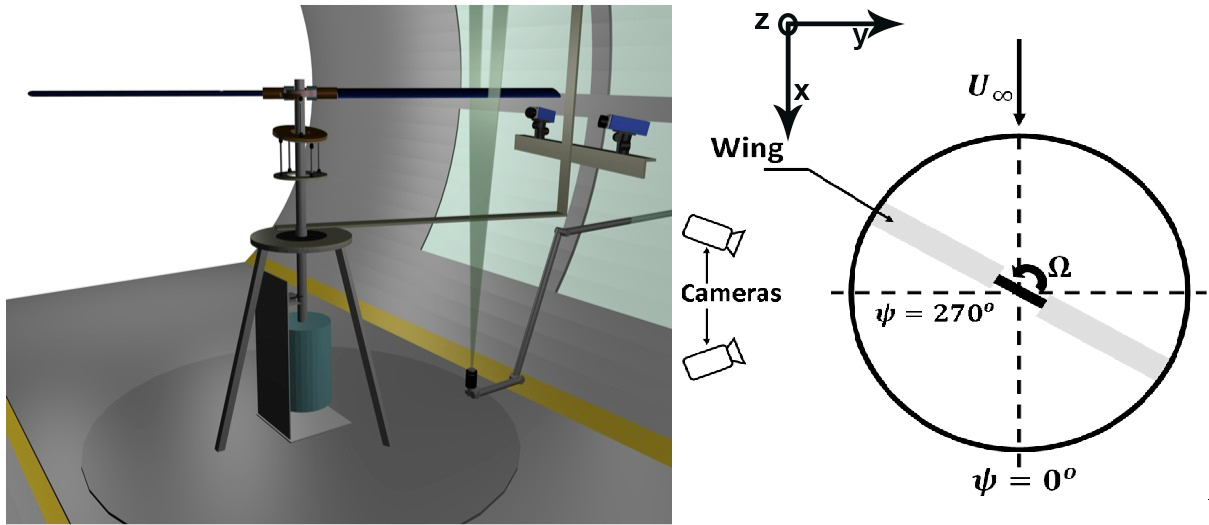}
	\caption{Reverse flow experimental setup }
	\label{TunnelNew}
\end{figure}

\begin{table}[!ht]
	\centering
	\caption{Rotor specifications}
	\begin{tabular}{ |c||c|c| }
		\hline
		Description & Value & Units \\
		\hline
		Blade mass & 1.747 & kg \\
		Blade span & 0.622 & m \\
		Blade chord & 0.178 & m \\
		Blade aspect ratio & 3.49 & - \\
		Disk radius & 0.889 & m \\
		Solidity & 0.089 & - \\
		Precone & 1.6 & degrees \\
		Max. collective & 10 & degrees \\
		Cyclic & 6.7 - 9.0 & degrees \\
		Motor & 3.73 & kW \\
		Height & 1.575 & m \\
		\hline
	\end{tabular}
	
	\label{Tab:RotorSpecs}
\end{table}

\section{Scope of this study}
The present work focuses on the mechanism of reverse flow on a rotor blade at high advance ratios. The paper delineates the evolution and dissipation phases of the sharp edge vortex based on the stereo particle image velocimetry (SPIV) measurements. The  characteristics of the SEV like convection speed, core size, and the velocity profile are studied. The morphological similarities of the SEV with the leading edge vortex on a delta wing at high angles of attack is described, and it proposes vortex aerodynamics as a key factor in this flow-field. 

\section{Experimental Method}

\subsection{Experimental setup}

The experiments were conducted in the John Harper 2.13 m x 2.74 m low speed wind tunnel at the Georgia Institute of Technology. It is a closed circuit wind tunnel built in the year 1920 and is located in the Guggenheim School of Aerospace Engineering.The rotors used in the rotating experiments were composed of two untwisted, rectangular, NACA 0013 rotor blades (identical to and including the one used in the static experiments discussed by Raghav \textit{et al.} \cite{4RaghavPhD2014}), attached to a teetering rotor hub actuated by a 3.73 kW motor. Figure \ref{TunnelNew} shows the rotor setup with the laser sheet and SPIV used in this work. The collective pitch angle was set at $7^{\circ}$ and the longitudinal cyclic angle was set at $8^{\circ}$, which results in a $15^\circ$ pitch at $\psi$ = $270^\circ$. These pitch angles created a tip path plane tilting forward, thus simulating the forward flight. And higher pitch angles resulted in rotor vibrations resulting from stall phenomenon. There was no lateral cyclic in these experiments. A detailed analysis on the blade surface roughness is described in our prior work \cite{4RaghavPhD2014}. The rotor specifications are shown in Table \ref{Tab:RotorSpecs}. 

\subsection{Flow and Test conditions}
The phase-locked SPIV measurements on the rotating rotor blade were acquired at advance ratios of $\mu$ = 0.7, 0.85, 1.0 and a rotor angular velocity of $\Omega = 20.94$ rad/s (200 RPM). Measurements were gathered at radial locations of r/R = 0.4, 0.5, 0.514, 0.6, and 0.7 and at azimuthal angles $\psi$ = $240^{\circ}$, $270^{\circ}$ and $300^{\circ}$. The azimuth angle of 240 degrees covered additional refined datasets at r/R = 0.35, 0.45, 0.5, 0.55, and 0.65. Table \ref{Tab:TestMatrixSPIV} summarizes the test cases for the SPIV measurements.  

\begin{table}[!t]
	\centering
	\caption{Test matrix for SPIV measurements}
	\begin{tabular}{ |c||c|c| }
		\hline
		Azimuth ($\Psi$)  & Advance ratio $\mu$  & r/R \\
		\hline
		$240^{\circ}$& 0.7, 0.85, 1.0 & 0.35, 0.4, 0.45, 0.5,  \\
		             &                & 0.514, 0.55, 0.6, 0.65, 0.68\\
		$270^{\circ}$& 0.7, 0.85, 1.0 & 0.4, 0.5, 0.514, 0.6, 0.68 \\
		$290^{\circ}$& 0.7, 0.85, 1.0 & 0.4, 0.5, 0.514, 0.6, 0.68 \\
		$300^{\circ}$& 0.7, 0.85, 1.0 & 0.4, 0.5, 0.514, 0.6, 0.68 \\
		\hline
	\end{tabular}
	
	\label{Tab:TestMatrixSPIV}
\end{table}

\begin{table}[!ht]
	\centering
	\caption{Uncertainty estimates in SPIV measurements}
	\begin{tabular}{ |c||c| }
		\hline
		Parameter  & Error\\
		\hline
		In-plane random error & 0.088 - 0.281 pixels\\
		In-plane velocity error($\varepsilon_u$, $\varepsilon_v$ )& 0.006 m/s - 0.02 m/s \\
		Out-of plane velocity error ($\varepsilon_w$) & 0.028 m/s - 0.281 m/s \\
		Total measurement error & 0.75\% - 3.96\% \\
		\hline
	\end{tabular}
	
	\label{Tab:UncertaintyEstimates}
\end{table}

\subsection{Measurement uncertainties}
The uncertainty in instantaneous velocity measurements is computed using methods described in \cite{raffel2004recording} and \cite{raffel2013particle}. The out of plane component error estimation  due to the relative angle of the cameras is discussed in \cite{lawson1997three}. For the camera angles used in these experiments the RMS error for the out-of-plane component of displacement is between 2.8\% - 6\% for a laser light sheet thickness of 3-4 mm and a camera magnification of 1/16. The error in out-of-plane component of velocity varies between 0.74\% to 3.95\%. This amounts to an absolute error of 0.028 m/s to 0.281 m/s based on maximum out of plane velocity component. The in-plane velocity measurement error was estimated to be 0.06\% to 0.2\%. This amounted to an absolute error of 0.006 m/s and 0.02 m/s based on maximum in-plane velocity component. The total velocity measurement error is computed as a percentage by computing the magnitude of all the measurement errors and it amounts to 0.75\% to 3.96\%. 
The uncertainty estimates in the SPIV measurements are summarized in Table \ref{Tab:UncertaintyEstimates}

\section{Results and Discussion}
\subsection{Azimuthal evolution of the SEV}

\begin{figure}[!t]
\center
\includegraphics[width =0.9\columnwidth]{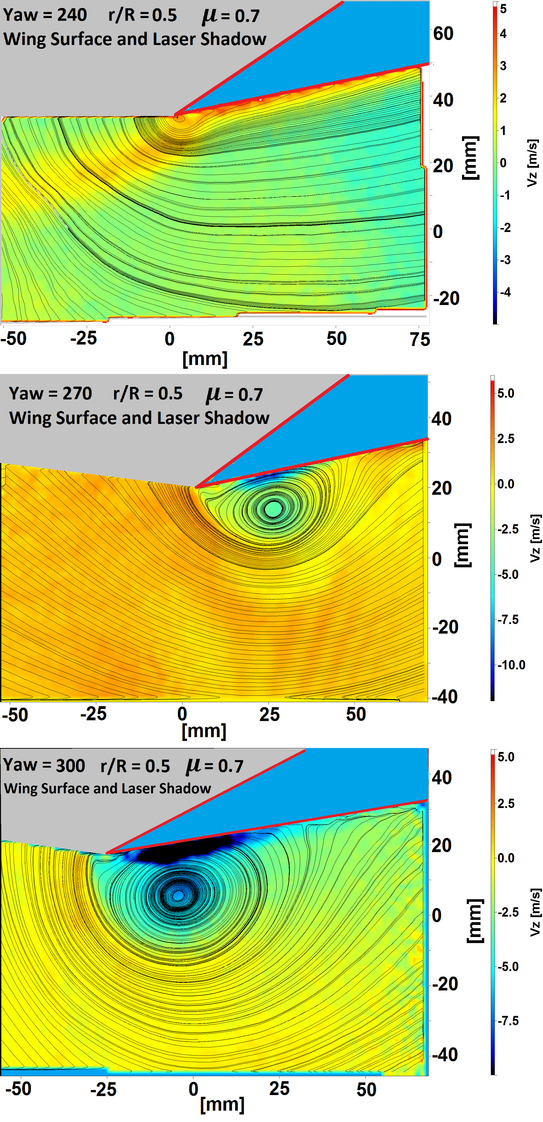}
\caption{Azimuthal evolution of SEV, $\mu = 0.7$ }
\label{AzimuthVariation1}
\end{figure}

At an azimuth of 240 degrees, the curvature of streamlines depicted in Figure \ref{AzimuthVariation1} shows an incipient SEV. The attached coherent vortex is observed very close to the surface originating from the sharp edge.  As the rotor blade progresses azimuthally, the SEV becomes stronger, larger and detached from the surface, but still convecting along with the rotor blade. Beyond 270 degrees, the vortex becomes even larger and eventually diffuses into the freestream. Unlike the generally accepted hypothesis of vortex shedding from sharp corners, no such phenomenon was observed in the data sets. All phase-locked SPIV data sets showed a coherent tight vortex with a small core size at the sharp edge convecting with the blade. The presence of the SEV well ahead of 270 degrees azimuth invalidates the reverse dynamic stall hypothesis wherein the presence of a vortex is related to the perturbations at 270 degrees.

At 240 degrees, the near-farfield flow and the vortical flow are directed radially inwards, towards the rotor hub. At 270 degrees azimuth, the near-farfield flow is directed radially outboard and the vortical flow still directed inwards. At 270 degrees, the rotor blade is oriented perpendicular to the free stream and there is no freestream component in the radial direction. Consequently, the near-farfield flow at 270 degrees is primarily due to centrifugal forces. The radially inward vortical flow is sufficiently strong to counteract the centrifugal stresses. At 300 degrees, the vortical flow is still directed radially inwards but diffused in nature. In this case, the freestream component is directed radially outwards which further strengthens the centrifugal forces that counteract the vortical flow. This interaction with the previously formed vortex resulting in a diffused vortex is termed as ``history effects".      

\begin{figure}[!t]
\center
\includegraphics[width =0.9\columnwidth]{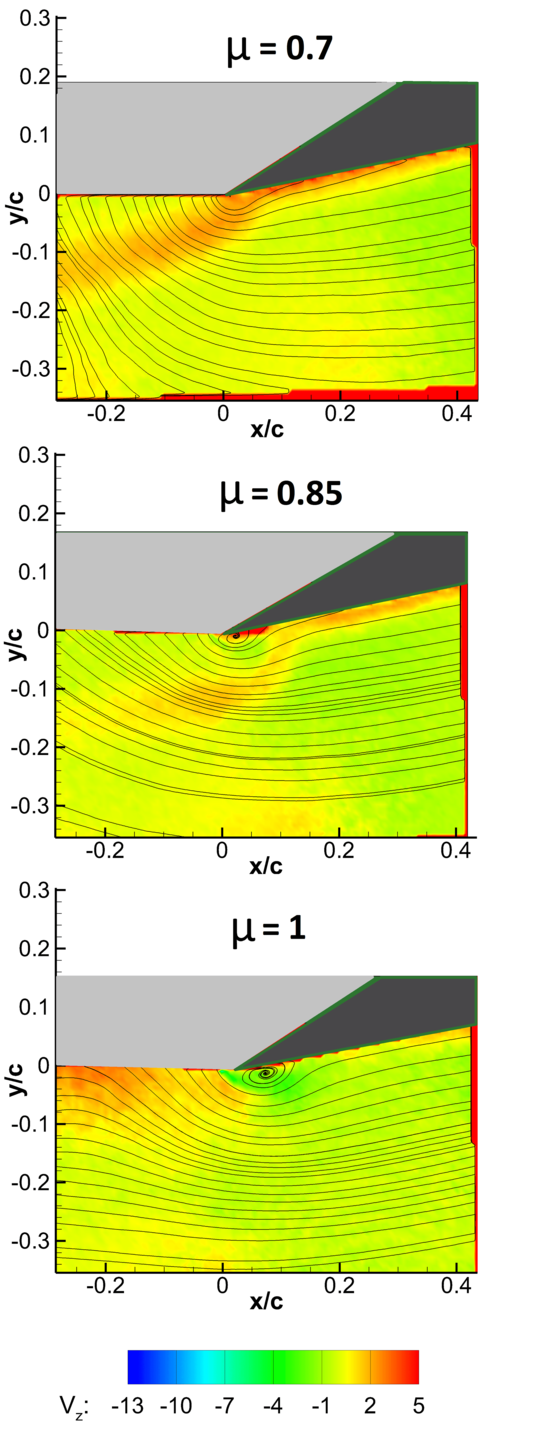}
\caption{Effects of advance ratio at r/R = 0.5, azimuth 240 degrees}
\label{Az240_AdvRatio}
\end{figure}

\subsection{Advance ratio effects on the SEV}

At azimuth 240 degrees, the increase in vortex strength with the increase in advance ratio is apparent from Figure \ref{Az240_AdvRatio}. The curvature of streamlines become more prominent at higher advance ratios. This is primarily due
to the increase in reverse flow with advance ratio. High reverse flow near the root causes a tight and strong vortex with a small core size very close to the sharp edge. There is no vortex shedding observed at this azimuth. Again, the development of the vortex at higher advance ratio is associated with a negative radial flow (towards the rotor hub). However, before the separation of the shear layer at the sharp edge, we see a predominant positive radial velocity (towards the blade tip). This positive radial velocity reverses direction right at/after the separation of the shear layer, leading to the formation of the vortex.

At 240 degrees, the strength of the vortex is evidently larger at higher advance ratios due to higher streamline curvatures. And at 270 degrees azimuth (not shown), the vortex core is larger at lower advance ratios. Further, the vortex core is detached from the surface at lower advance ratios. And a similar behavior of the detached vortex is observed at azimuth 300 degrees. In addition, a strong inward directed radial flow is prominently seen near the surface, but away from the vortex center. This is in contrast to the observed radial flow in the core of the vortex at 270 degrees azimuth. The detached vortex observed at azimuth angles of 270 and 300 degrees still convects along with the rotor blade and does not diffuse/convect with the freestream.

\subsection{Radial variation of the SEV}
\begin{figure}[!t]
\center
\includegraphics[width =1\columnwidth]{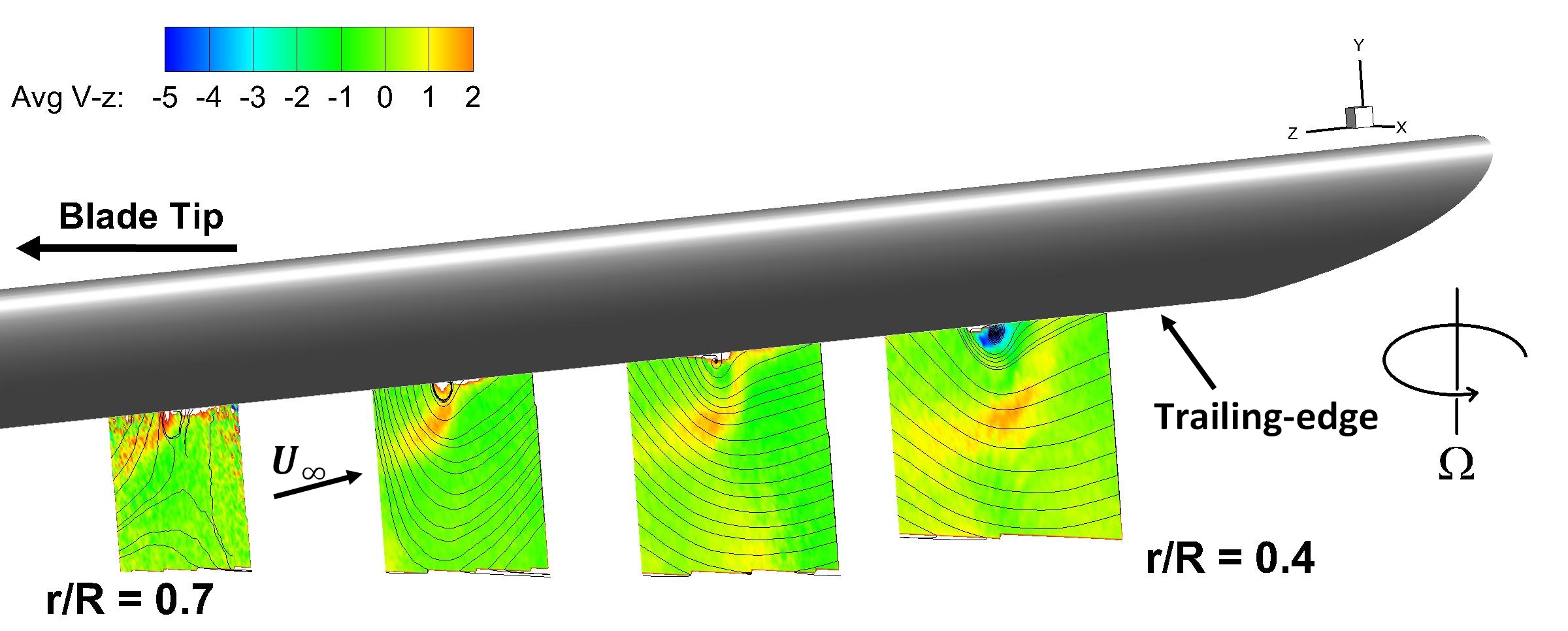}
\caption{Radial variation of SEV at $\Psi=240^{\circ}$, $\mu=0.85$}
\label{240az_variation_rad}

\includegraphics[width =1\columnwidth]{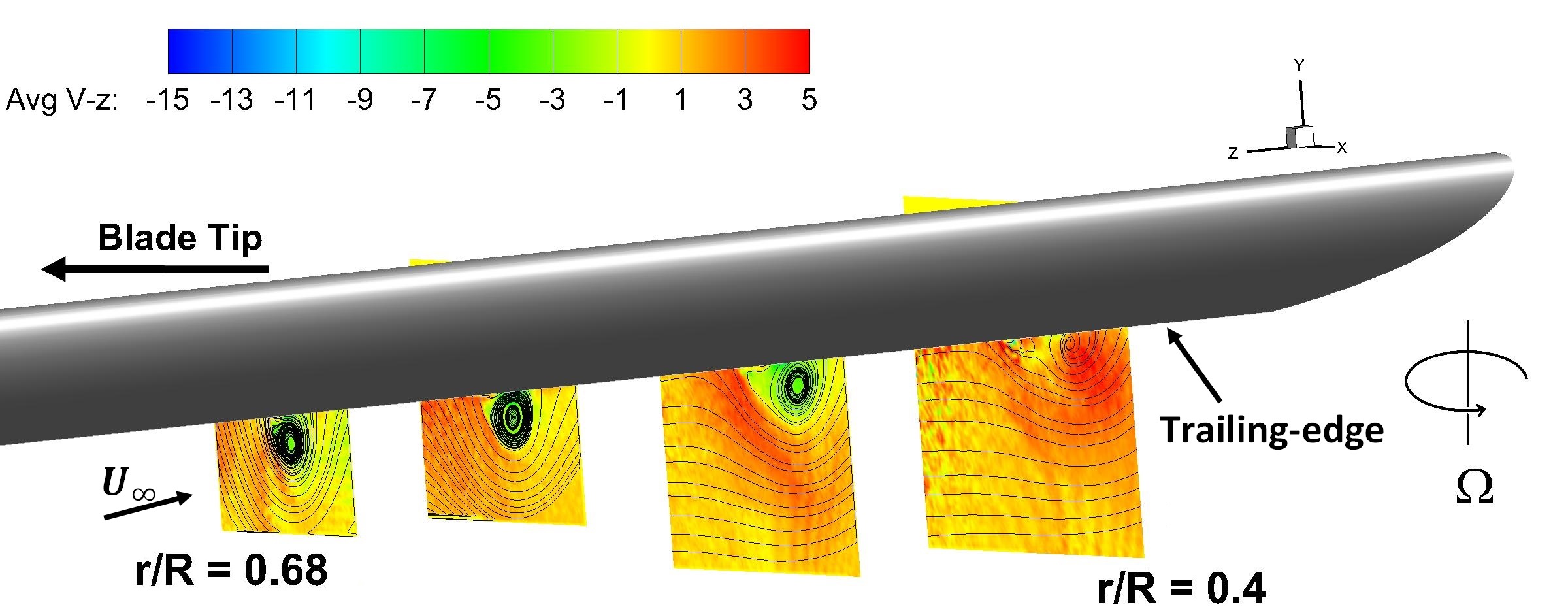}
\caption{Radial variation of SEV at $\Psi=270^{\circ}$, $\mu=0.85$}
\label{270az_variation_rad}

\includegraphics[width =1\columnwidth]{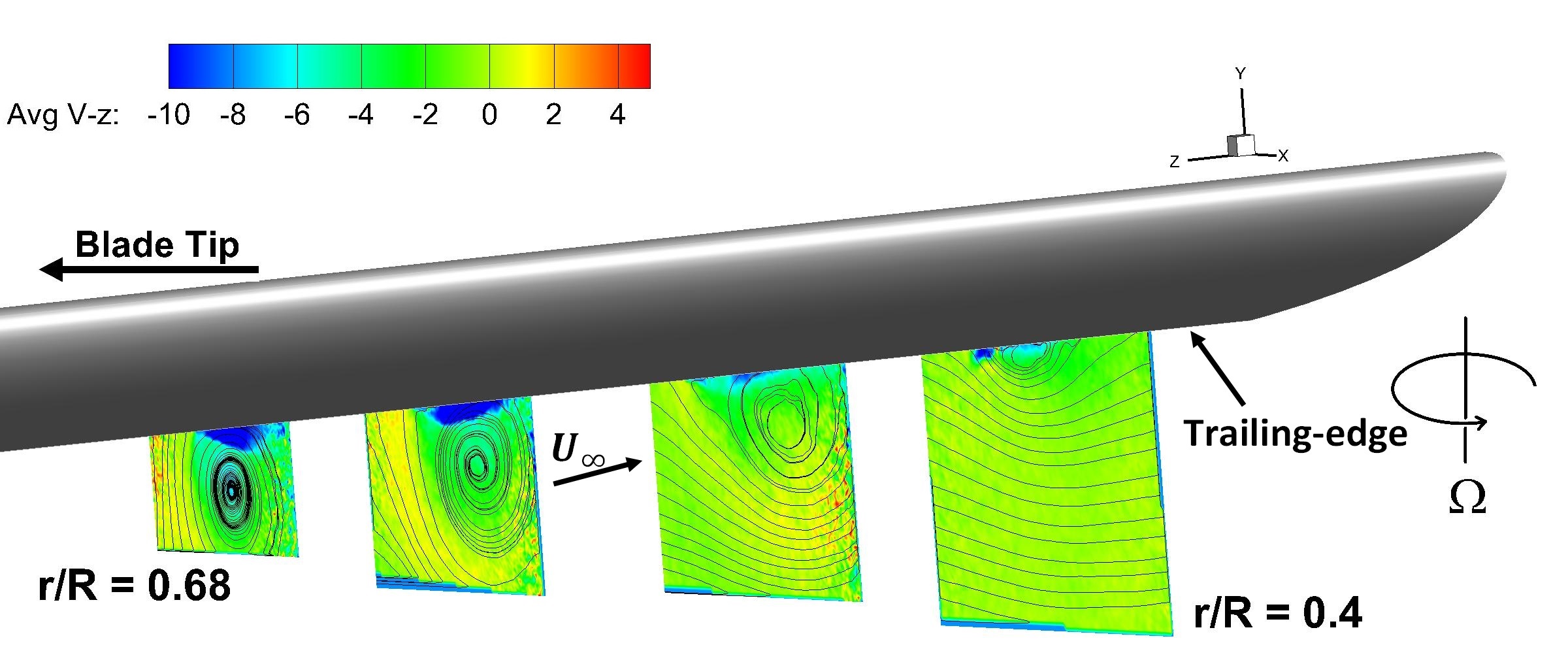}
\caption{Radial variation of SEV at  $\Psi=300^{\circ}$, $\mu=0.85$}
\label{300az_variation_rad}
\end{figure}

Figure \ref{240az_variation_rad} shows an incipient vortex at r/R of 0.6 that grows radially inwards. There is no evidence of the vortex at r/R of 0.7 which is outside the reverse flow region for an advance ratio of 0.85. The start of the reverse flow region is the source of vortex generation. 

At 270 degrees, the vortex originates at a further outboard location of r/R of 0.68. Figure \ref{270az_variation_rad} shows a smaller vortex core near the tip that becomes stronger and tighter as it moves radially inward. The directionality of the vortex flow can be clearly distinguished by the outward flow in the near-farfield, thereby validating a strong helical vortex. Also, it is evident that the flow inside the vortex is directed towards the root while counteracting the centrifugal forces at the surface. \index{helical structure}   

At 300 degrees, shown in Figure \ref{300az_variation_rad}, the vortex is detached from the surface and shows the signs of a burst vortex with the solid core extending all the way to the outer diameter. The shear layer of feeding vorticity into the vortex shows an outward directed flow. The inward directed flow is observed much closer to the surface and away from the vortex core. A clear distinction of the vortex origin at a particular radial location is questionable at this stage due to the history effects. The aggravating centrifugal forces on a previously formed vortex need to be considered. In contrast to the 240 degrees case at r/R 0.68, which clearly falls outside the reverse flow region, the presence of the vortex at 300 degrees azimuth clearly indicates a previously formed convected vortex rather than an incipient vortex. 

\subsection{Vortex convection}

\begin{figure}[!ht]
\center
\includegraphics[width =1.0\columnwidth]{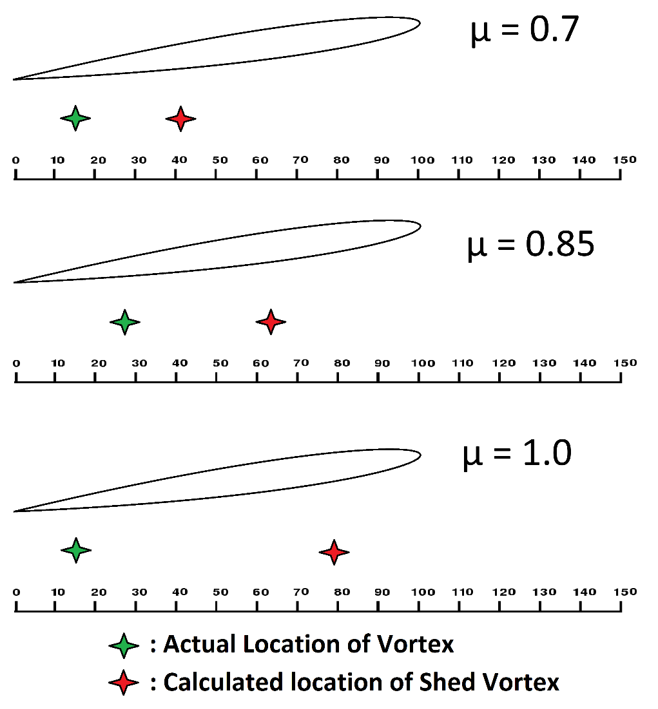}
\caption{Relative position of SEV vs shed vortex at $\Psi=300^{\circ}$, $r/R=0.5$}
\label{VortexConvection}
\end{figure}

From the understanding of a convecting vortex, the convection speed of a free vortex in a freestream lies between 0.4-0.6 $U_{\infty}$. A validation study was performed to obtain the convective speeds and the relative position for a free shed vortex, if this were to be the case. A vortex was presumed to be shed at r/R of 0.5 at an azimuth of 270 degrees. The convection times for different advance ratios for a supposedly shed vortex were found to be of the order of 18-25 ms.The relative position of a shed vortex with respect to the blade at an azimuth angle of 300 degrees is thus shown in Figure \ref{VortexConvection}.

The experimentally observed SEV in the velocity fields did not show such a behavior, as a shed vortex would have convected further downstream of the sharp edge. In contrast, the SEV was observed to be at a distance of 15\%c to 28\%c from the sharp edge. The presence of the SEV near the sharp edge even at 300 degrees azimuth validates our hypothesis that the SEV truly convects with the rotor blade. 

\begin{figure}[!ht]
	\includegraphics[width=1\linewidth]{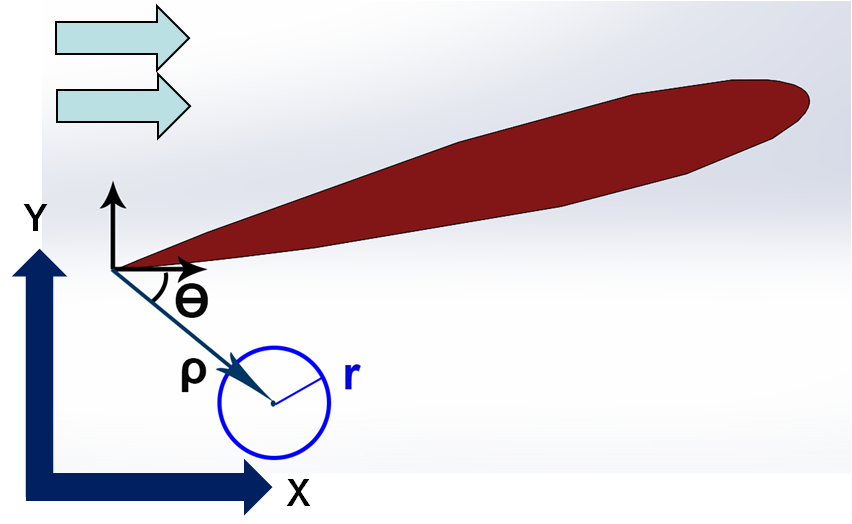} 
	\caption{Position of SEV from the sharp edge}
	\label{VortexSchematic}
	
\end{figure}

The relative position of the SEV and the sharp edge vary as one moves radially inwards for different advance ratios and azimuths. The vortex center is identified algorithmically based on the vortex core size. The vortex position is characterized by the orientation angle $\theta$ of the position vector, with its magnitude denoted as $\rho$. The origin of the position vector is located at the sharp edge for a given radial location as shown in Figure \ref{VortexSchematic}. 

Tables \ref{Tab:SevPositionAz240} to \ref{Tab:SevPositionAz300} shows the characterization of the SEV. At azimuth 240 degrees the SEV is very close to the sharp edge and the $\rho/c$ varies between 0.014 to 0.072. At lower advance ratio, the SEV is observed to be closer ($\rho/c = 0.014$) to the sharp edge than at higher advance ratio ($\rho/c = 0.072$). The position of the SEV at inboard locations is slightly higher than at the outboard locations. Similar behavior is observed at azimuth 270 and 300 degrees with a greater separation distance from the sharp edge at higher azimuth angles.

\begin{table}[!ht]
	\centering
	\caption{Position of SEV at $\Psi = 240^{\circ}$}
	\begin{tabular}{ |c||c|c|c|c|c|c|c|c|c| }
		\hline
		\multicolumn{1}{|c||}{$r/R$} & \multicolumn{3}{c}{$\mu = 0.7$} & \multicolumn{3}{|c}{$\mu = 0.85$} & \multicolumn{3}{|c|}{$\mu = 1.0$} \\
		\hline
		- & $\rho/c$ & ${\theta}^\circ $ & $r_{\circ}/c$ & $\rho/c$ & ${\theta}^\circ  $ & $r_{\circ}/c$ & $\rho/c$ & ${\theta}^\circ $ & $r_{\circ}/c$ \\
		  & & & (\%)& & & (\%)& & & (\%)\\
		\hline
		0.35 & 0.028 & -28 & 5.61 & 0.056 & -9 & 6.50 & 0.065 & -3 & 6.35 \\
		0.40 & 0.010 & -41 & 4.43 & 0.037 & -16& 4.84 & 0.072 & -1 & 3.49\\
		0.45 & 0.009 & -63 & 4.36 & 0.035 & -12& 5.26 & 0.047 & -9 & 5.00 \\
		0.50 & 0.019 & -9  & 3.30 & 0.210 & -3 & 3.55 & 0.232 & -4 & 6.12\\
		0.514& 0.014 & -18  & 3.86& 0.011 & -21& 4.08 & 0.047 & -15 & 5.11\\
		0.55 & 0.013 & -8  & 2.52 & 0.011 & -28& 4.06 & 0.030 & -10 & 4.51\\
		0.60 & 0.019 & -4  & 0.88 & 0.221 & -2 & 2.73 & 0.031 & -82 & 3.621\\
		0.65 & 0.041 & -6  & 0.65 & 0.073 & -19& 0.27 & 0.022 & -77 & 2.984\\
		0.68 & 0.014 & -17 & 1.09 & 0.031 & -45& 2.45 & 0.038 &
		-76 & 2.789\\
		\hline
	\end{tabular}
	
	\label{Tab:SevPositionAz240}
\end{table}

\begin{table}[!ht]
	\centering
	\caption{Position of SEV at $\Psi = 270^{\circ}$}
	\begin{tabular}{ |c||c|c|c|c|c|c|c|c|c| }
		\hline
		\multicolumn{1}{|c||}{$r/R$} & \multicolumn{3}{c}{$\mu = 0.7$} & \multicolumn{3}{|c}{$\mu = 0.85$} & \multicolumn{3}{|c|}{$\mu = 1.0$} \\
		\hline
		- & $\rho/c$ & ${\theta}^\circ$ & $r_{\circ}/c$ & $\rho/c$ & ${\theta}^\circ $ & $r_{\circ}/c$ & $\rho/c$ & ${\theta}^\circ $ & $r_{\circ}/c$ \\
		 & & & (\%)& & & (\%)& & & (\%)\\
		\hline
		0.40 & 0.198 & -12 & 10.39 & 0.233 & -2	& 7.14 & 0.178 & -1.9 & 7.46\\
		0.50 & 0.139 & -18  & 9.22 & 0.184 & -11 & 7.84 & 0.103 & -1 & 4.15\\
		0.514& 0.125 & -18  & 9.24& 0.175 & -8 & 8.56 & 0.119 & -4 & 6.53\\
		0.60 & 0.064 & -36  & 6.393 & 0.126 & -14 & 8.13 & 0.087 & -5 & 6.78\\
		0.68 & 0.006 & -38 & 5.95 & 0.069 & -19& 6.12 & 0.048 &
		-19 & 2.65\\
		\hline
	\end{tabular}
	
	\label{Tab:SevPositionAz270}
\end{table}

\begin{table}[!ht]
	\centering
	\caption{Position of SEV at $\Psi = 300^{\circ}$}
	\begin{tabular}{ |c||c|c|c|c|c|c|c|c|c| }
		\hline
		\multicolumn{1}{|c||}{$r/R$} & \multicolumn{3}{c}{$\mu = 0.7$} & \multicolumn{3}{|c}{$\mu = 0.85$} & \multicolumn{3}{|c|}{$\mu = 1.0$} \\
		\hline
		- & $\rho/c$ & ${\theta}^\circ $ & $r_{\circ}/c$ & $\rho/c$ & ${\theta}^\circ $ & $r_{\circ}/c$ & $\rho/c$ & ${\theta}^\circ $ & $r_{\circ}/c$ \\
		 & & & (\%)& & & (\%)& & & (\%)\\
		\hline
		0.40 & 0.245 & -36 & 12.018 & 0.139 & -25	& 10.45 & 0.011 & -31 & 9.22\\
		0.50 & 0.185 & -29  & 11.110 & 0.077 & -18 & 7.48 & 0.118 & -25 & 9.29\\
		0.514& 0.151 & -31  & 10.626 & 0.016 & -20 & 3.32 & 0.155 & -17 & 10.40\\
		0.60 & 0.138 & -53  & 9.772 & 0.231 & -27 & 10.14 & 0.182 & -19 & 8.64\\
		0.68 & 0.125 & -90 & 3.57 & 0.230 & -89& 8.14 & 0.180 &
		-89 & 6.81\\
		\hline
	\end{tabular}
	
	\label{Tab:SevPositionAz300}
\end{table}

\subsection{SEV convection speed}

The convection speed provides an insight on the relative position of the SEV with respect to the rotor blade. The convection speeds are essential to study the kinetic energy of the SEV in comparison with rotational energy at different advance ratios. The convection speeds ($\bar{U}_P$) are calculated as an average over the vortex area as shown in Equation \ref{ConvecSpeed}, wherein the subscript `P' denotes the vortex center.  

\begin{equation}\label{ConvecSpeed}
\bar{U}_P = \frac{1}{S}\int_{S}U dS
\end{equation}

\begin{figure}[!ht]
	\includegraphics[width=1\linewidth]{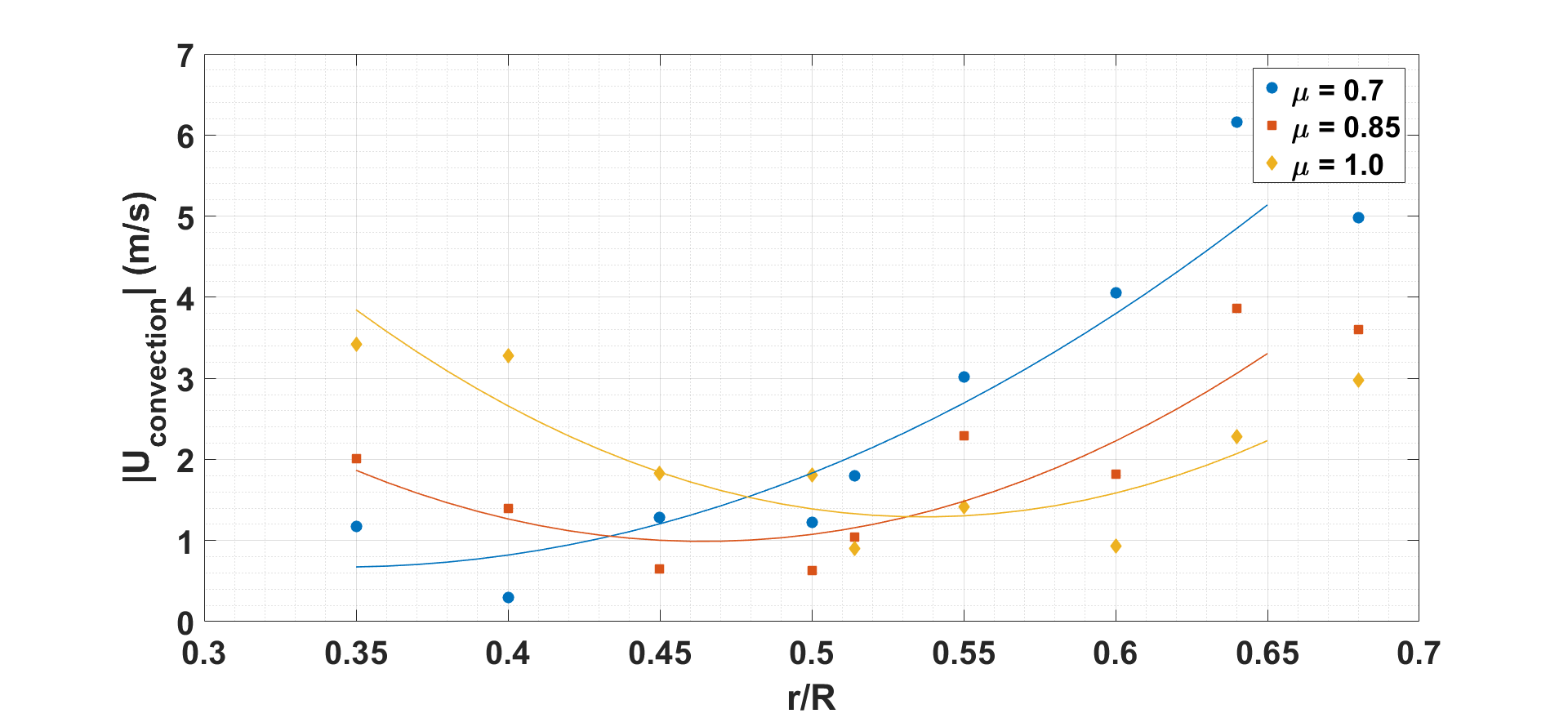} 
	\caption{Vortex convection speed for $\Psi = 240^\circ$}
	\label{ConvectionSpeed240}
	
	\includegraphics[width=1\linewidth]{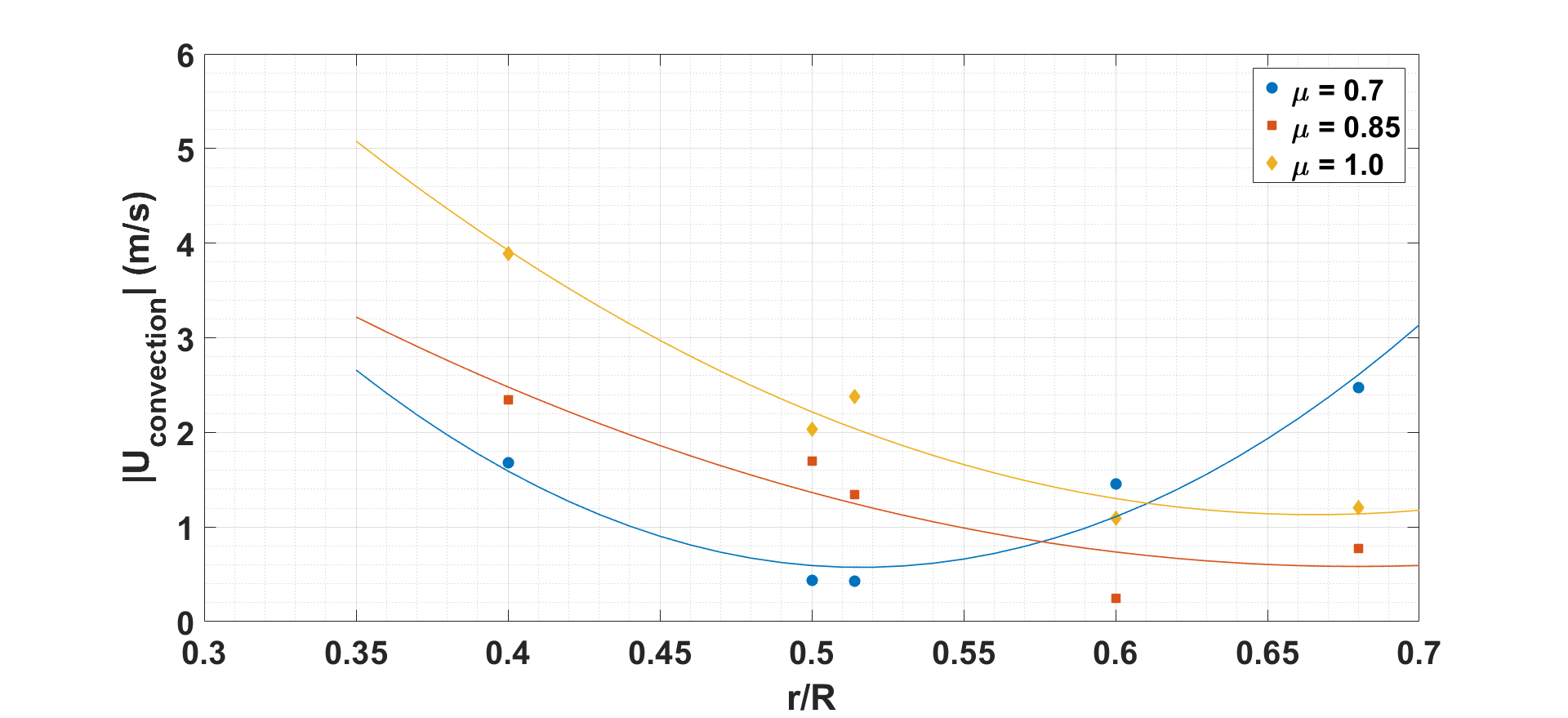}
	\caption{Vortex convection speed for $\Psi = 270^\circ$}
	\label{ConvectionSpeed270}
	
	\includegraphics[width=1\linewidth]{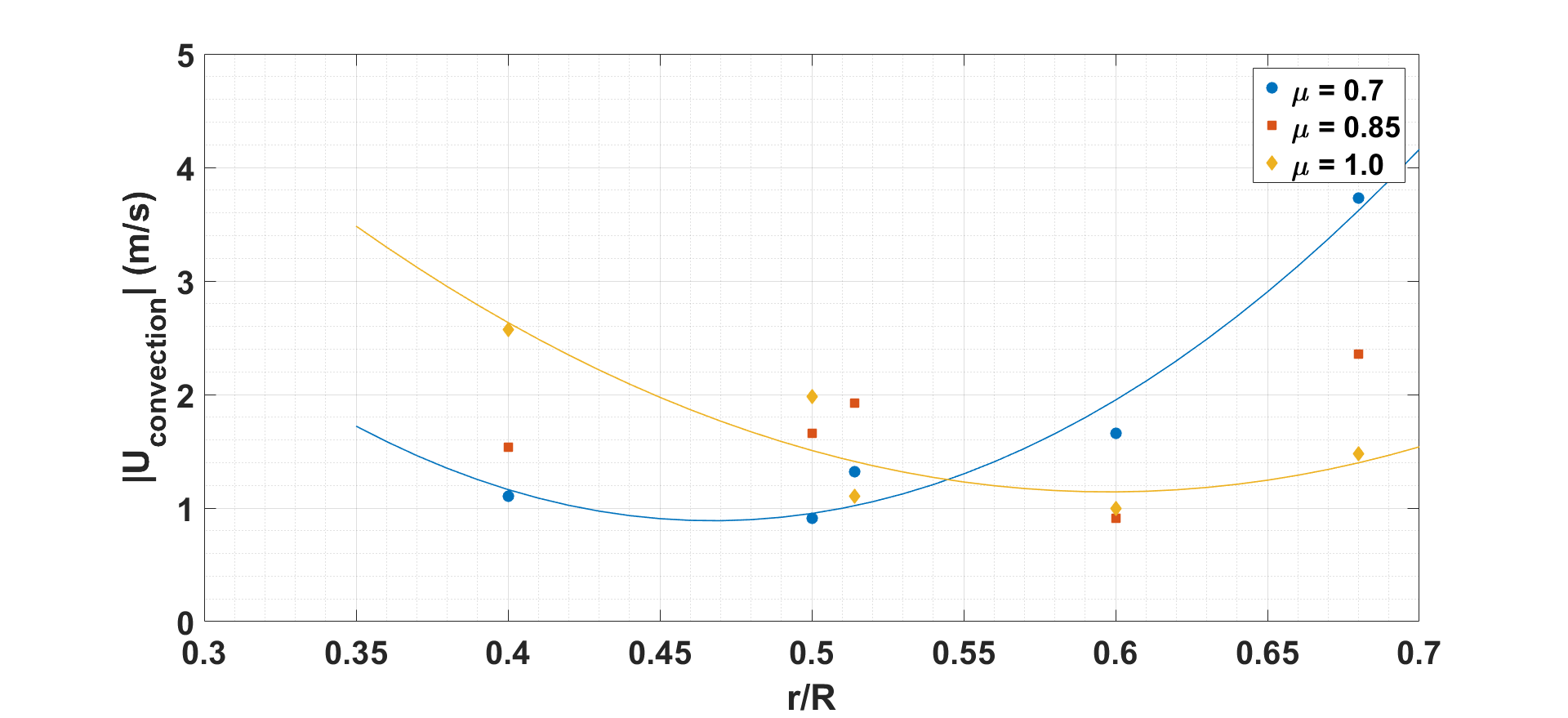}
	\caption{Vortex convection speed for $\Psi = 300^\circ$}
	\label{ConvectionSpeed300}
	
\end{figure}

The convection speed at 240 degrees azimuth shown in Figure \ref{ConvectionSpeed240} shows a parabolic behavior and the minimum shifts outboards with increase in advance ratio. At outboard radial locations, the forming vortex has lower rotational energy. With increase in advance ratio, this rotational energy increases with a proportional decrease in the kinetic energy drawn from the freestream. At an advance ratio of 0.7 the vortex convection speed monotonically decreases as one moves radially inwards. For higher advance ratios this trend stops at r/R of 0.5 and the convection speeds increase at the inboard radial stations.  

At azimuth 270 degrees shown in Figure \ref{ConvectionSpeed270}, the vortex convection speed decreases in the outboard radial locations; however at the advance ratio of 0.7 the convection speed still shows a parabolic behavior. The presence of a minimum in the convection speed r/R of 0.5 has direct correlation with the observed vorticity. The vorticity plots show a  distinct and perfectly circular SEV at this radial location, indicating the right amount of rotational energy required to sustain the coherent structure. At azimuth 300 degrees shown in Figure \ref{ConvectionSpeed300}, no particular behavior was observed at an advance ratio of 0.85, however the parabolic trend is consistent at advance ratios 0.7 and 1.0. In comparison with the convection speeds of a freely shed vortex ($0.4U_{\infty} - 0.6U_{\infty}$) discussed earlier, the convection speed of the SEV is observed to be $0.07U_{\infty} - 0.27U_{\infty}$.

\subsection{Vortex size}
\begin{figure}[!ht]
	\includegraphics[width=1\linewidth]{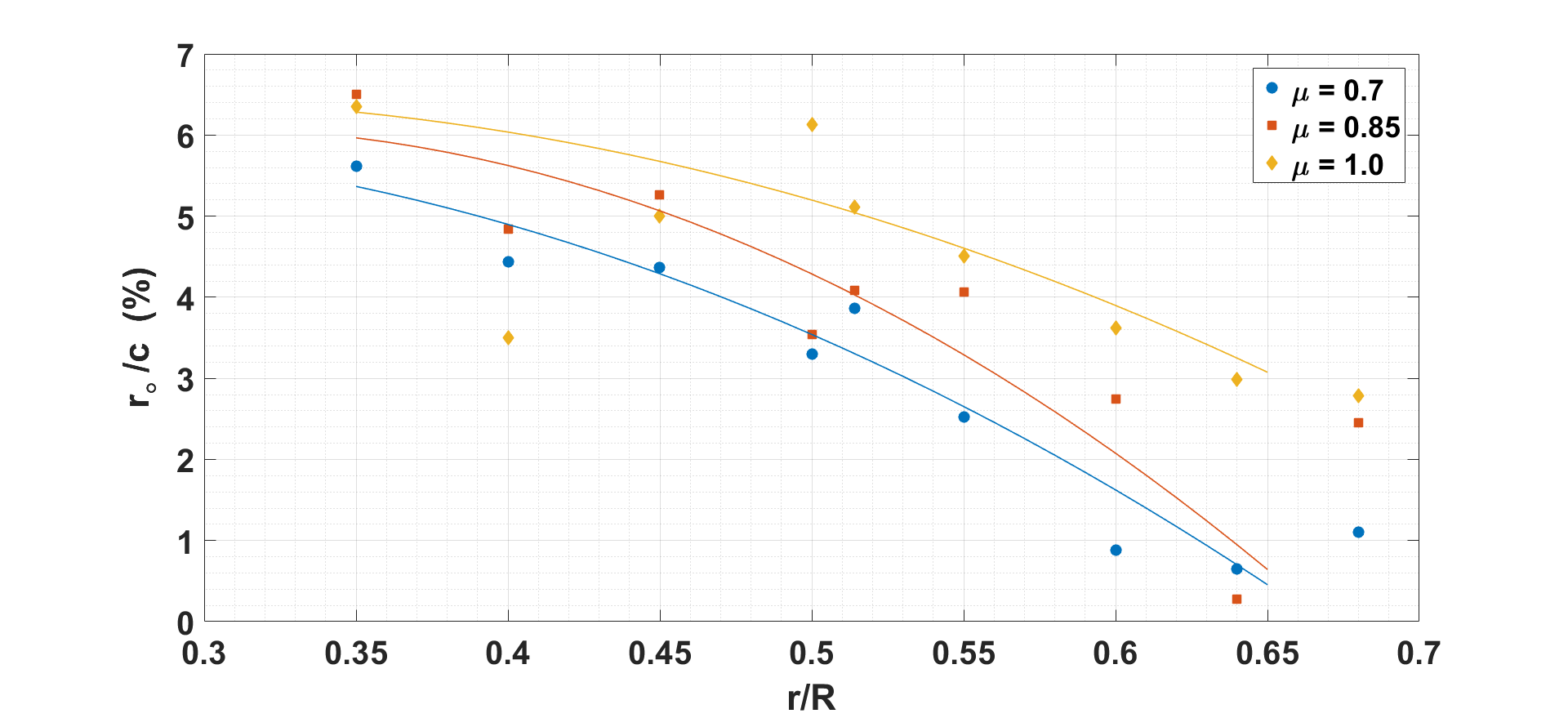} 
	\caption{Vortex core radius for $\Psi = 240^\circ$}
	\label{VortexRadius240}
	
	\includegraphics[width=1\linewidth]{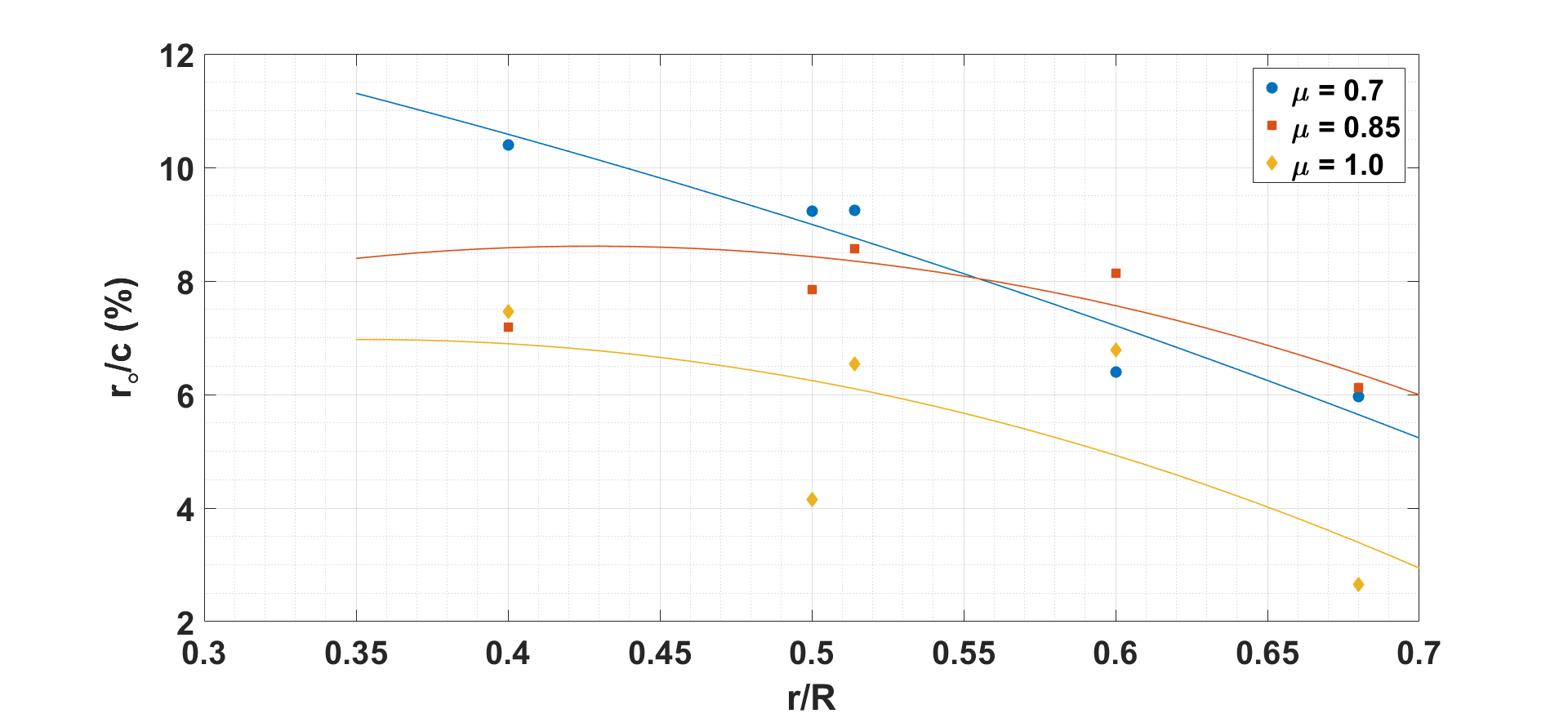}
	\caption{Vortex core radius for $\Psi = 270^\circ$}
	\label{VortexRadius270}
	
	\includegraphics[width=1\linewidth]{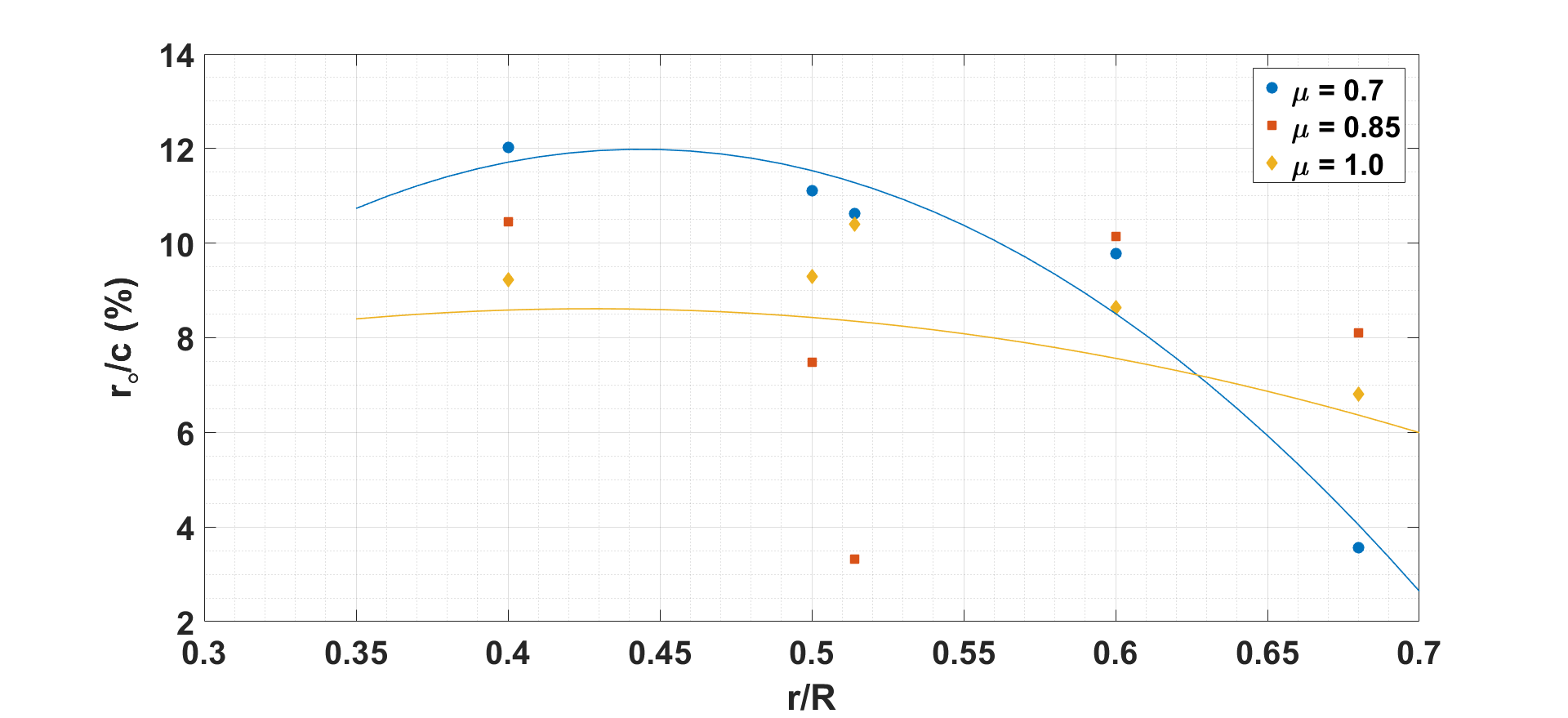}
	\caption{Vortex core radius for $\Psi = 300^\circ$}
	\label{VortexRadius300}
	
\end{figure}

The core size of vortex gives a measure of evolution of the SEV. The core size is defined by a region of linear velocity variation describing the extent of solid-body core rotation. The core radius is identified algorithmically by identifying the inflection point in the velocity profile measured across the diameter of the vortex. The algorithm is explained in detail by Michard \textit{etal}. \cite{graftieaux2001combining}. Figures \ref{VortexRadius240}, \ref{VortexRadius270} and \ref{VortexRadius300} show the size of the vortex core at different advance ratios and azimuths. 

At 240 degrees azimuth, the vortex size increases radially from 1\% of chord length at the start of the reverse flow region to 6.5\%. The SEV originates at a further outboard radial location at higher advance ratios as the extent of the reverse flow region increases. The size of the vortex core increases from 4\% to 11.5\% at 270 degrees azimuth. And at 300 degrees azimuth, the vortex core size is of similar size as it was at 270 degrees. The increase in the vortex core size from 240 degrees to 270 degrees shows the evolution of SEV. The stagnated growth beyond 270 degrees shows a quantitative measure of ``history effects". The 240 degrees azimuth shows a clear trend in the vortex core size with change in advance ratio. Such a behavior is inconsistent at other azimuths.

\subsection{Velocity profiles of the SEV}

A Galilean invariance method \cite{graftieaux2001combining}. 
denoted by $\Gamma_2(P)$, where the vortex structure is identified in an inertial reference frame is used to obtain the convection speed and the center of the vortex. The mean convection speed $\bar{U}_P$ of the vortex is subtracted from the observed velocity field in the rotor reference frame. The point of maximum correlation denotes the center of the vortex. The extent of the vortex core is determined by the position where $\Gamma_2(P) = 2/\pi$.   

\begin{figure}[!ht]
	\includegraphics[width=1\linewidth]{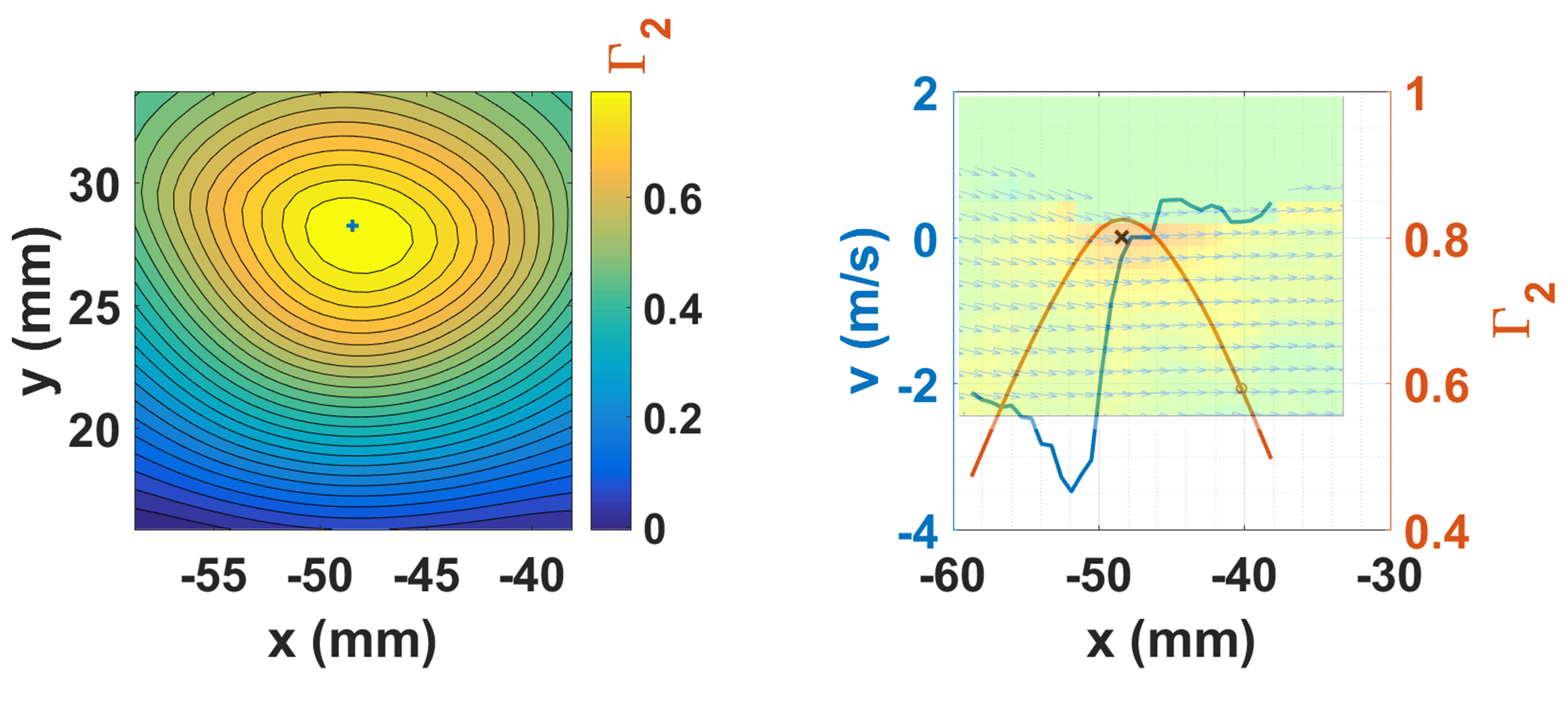} 
	\caption{Vortex velocity profile at $\Psi = 240^\circ$,$r/R = 0.45$,  $\mu = 0.7$}
	\label{az240rR0.45mu0.7ref47_p}

    \includegraphics[width=1\linewidth]{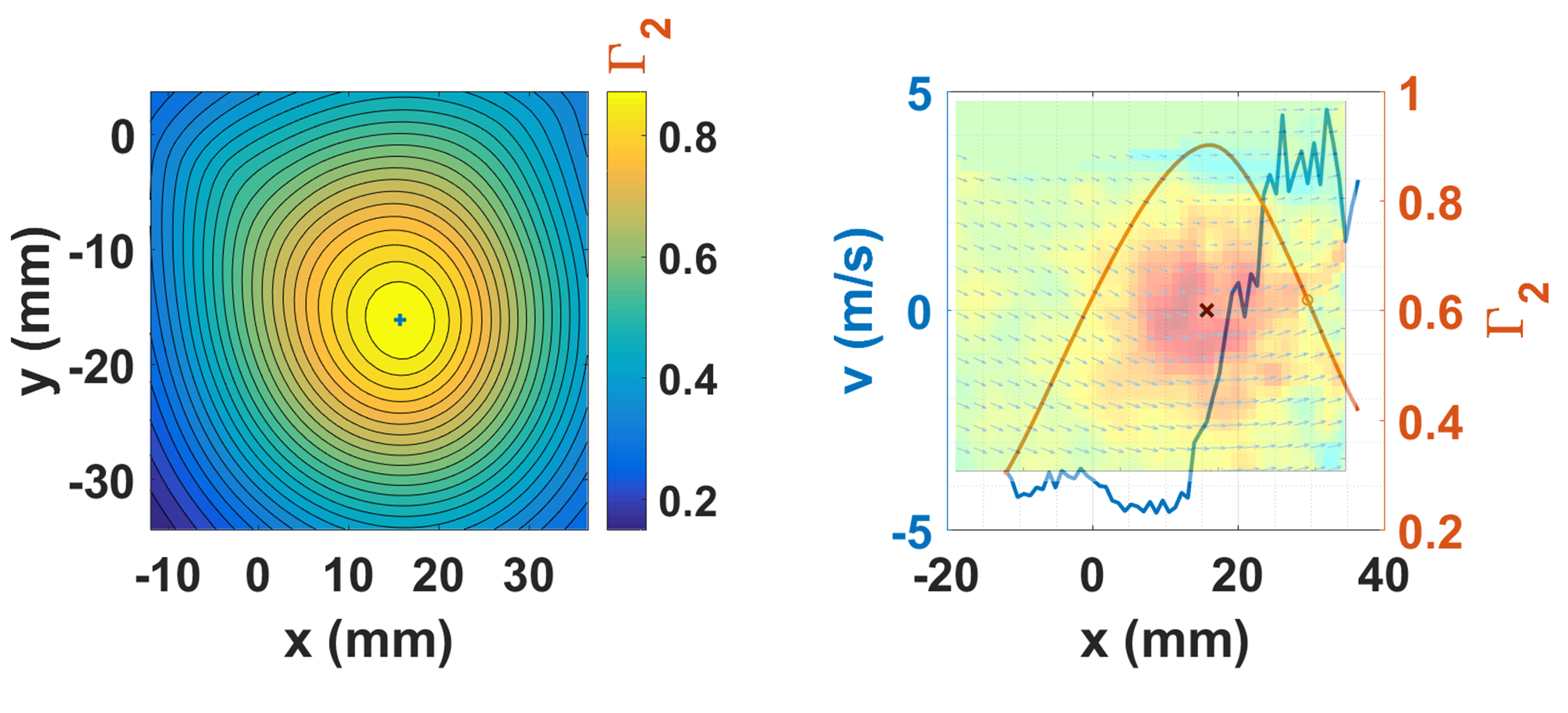}
    \caption{Vortex velocity profile at  $\Psi = 300^\circ$, $r/R = 0.6$, $\mu = 0.85$}
    \label{az300rR0.6mu0.85ref189_p}

\end{figure}

The velocity profile of a typical vortex comprises of two regions, a solid core rotation that has a linear velocity profile, and a irrotational region that is dictated by the $1/r^2$ rule for the velocity profile. Analogous to leading edge vortex on a delta wing, the SEV shows similar characteristics of velocity profile. At azimuth angles of 240 degrees when the rotor blade has a forward sweep, a higher relative angles of attack are observed across the span. The vorticity from the sharp edge is fed into the growing vortex while the axial pressure gradient in the core sustains the vortex by counteracting the centrifugal stresses. Figure \ref{az240rR0.45mu0.7ref47_p} shows the velocity profile of the vortex. The contour plot on the left shows the non dimensional values of $\Gamma_2(P)$ to determine the center, and it is also depicted by the red curve in the right image. The blue line overlaid on the vorticity contour shows the velocity profile. The short extant of the linear region in the solid rotation shows the existence of a very small vortex core. And outside the core, the velocity profile follows the $1/r^2$ rule. At azimuth angles of 300 degrees, the vortex core is stretched all the way to the outer extent of the vortex. As shown in Figure \ref{az300rR0.6mu0.85ref189_p}, the $1/r^2$ rule is not observed outside the vortex core. This shows the signs of a typical burst vortex that is commonly observed on the delta wings at very high angles of attack as the leading vortex disintegrates.    

\begin{equation}
\Gamma_2(P) = \frac{1}{N} \sum_{S}\frac{(PM\Lambda(U_M - \bar{U}_P)).\hat{z}}{||PM||\cdot||(U_M - \bar{U}_P)||}
\end{equation}

\begin{figure}
	\includegraphics[width=1\linewidth]{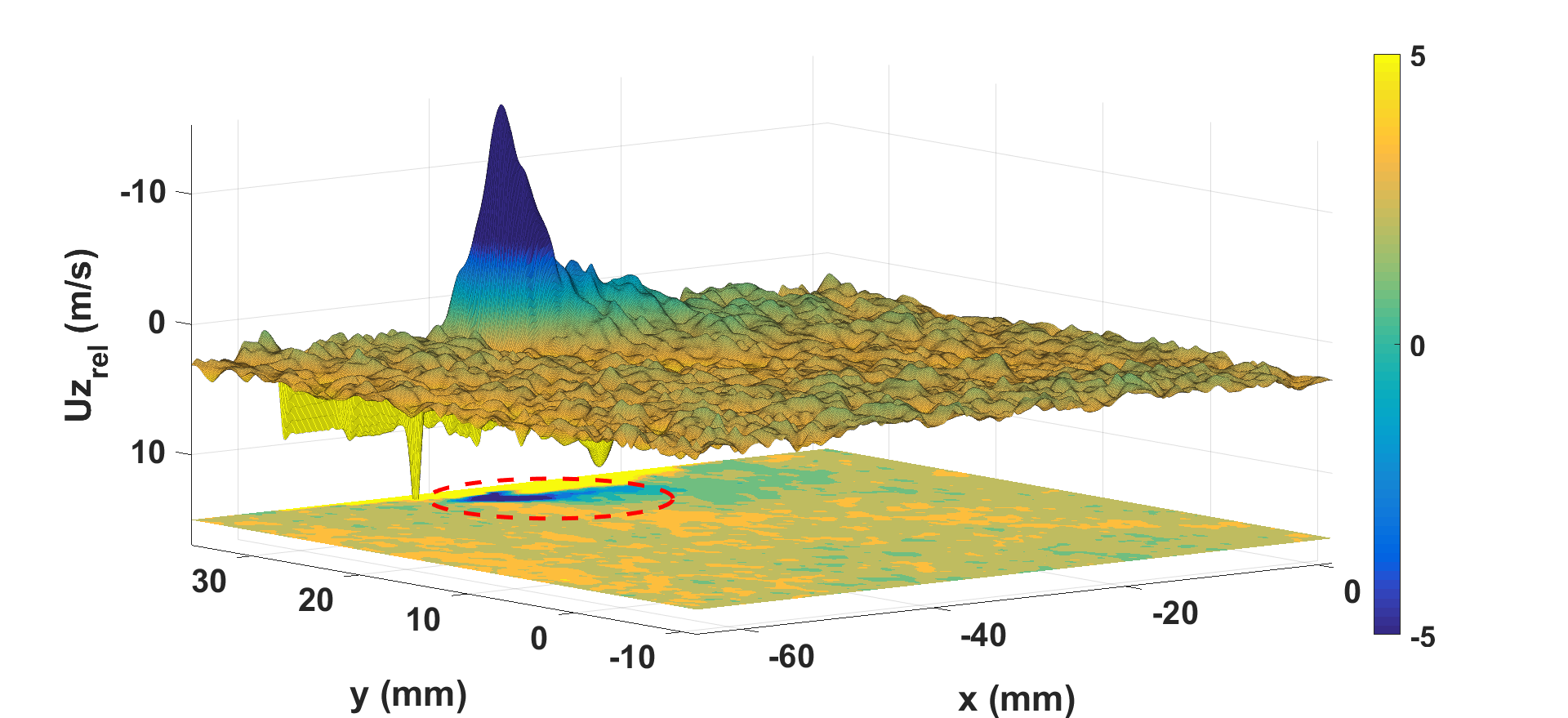}
	\caption{Axial velocity profile at  $\Psi = 240^\circ$, $r/R = 0.4$,  $\mu = 1.0$}
	\label{az240rR0.4mu1.0ref49_SEUzRel}
	
	\includegraphics[width=1\linewidth]{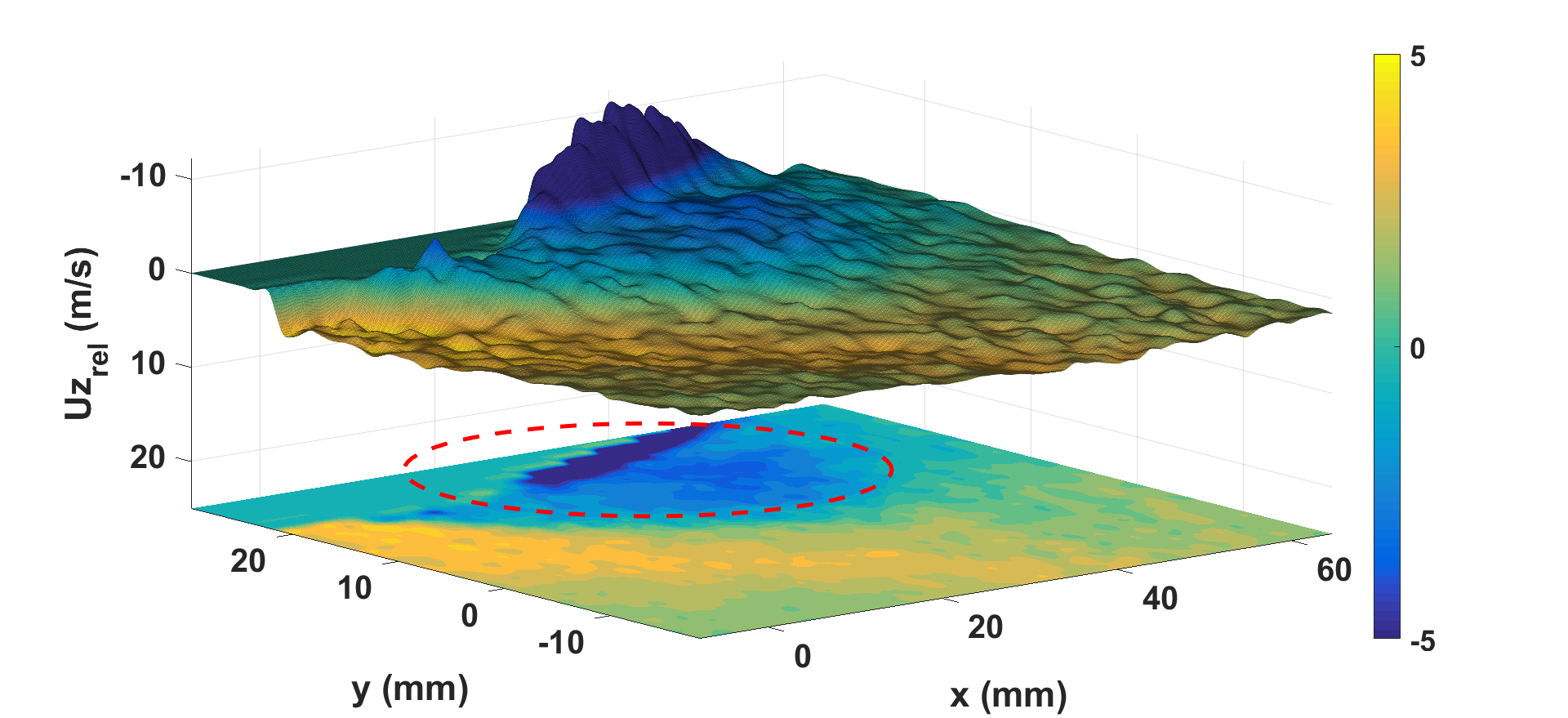}
	\caption{Axial velocity profile at  $\Psi = 270^\circ$, $r/R = 0.4$,  $\mu = 0.85$}
	\label{az270rR0.4mu0.85ref168UzRel}
	
		\includegraphics[width=1\linewidth]{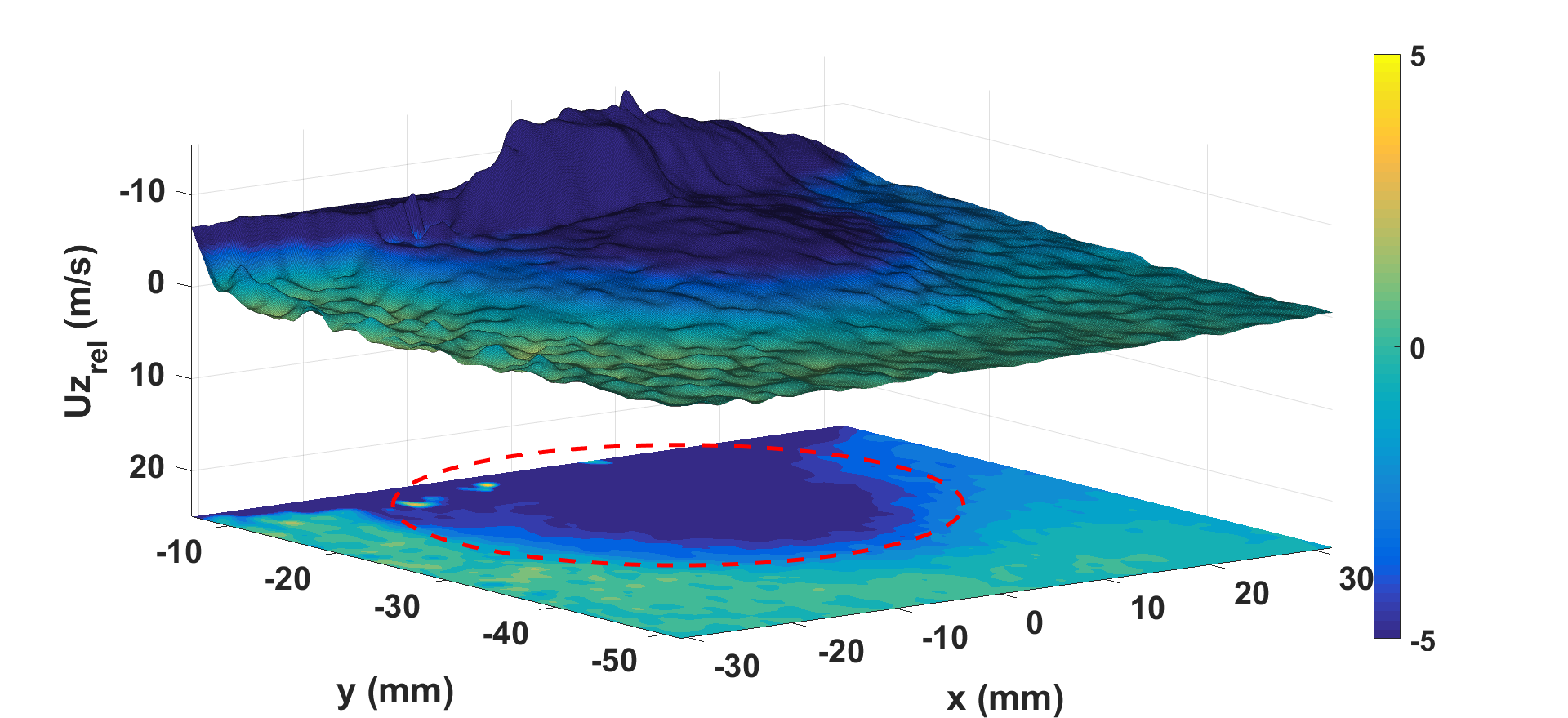} 
	\caption{Axial velocity profile at  $\Psi = 300^\circ$, $r/R = 0.514$, $\mu = 0.7$}
	\label{az300rR0.514mu0.7ref187.5UzRel}

\end{figure}

The axial velocity of the SEV is significant and the flow-field is highly three dimensional. The axial velocity profiles show the presence of a helical vortex with a strong axial component. The velocity surfaces are plotted for axial velocities ($w$) by subtracting the spanwise component of the freestream as shown in Equation \ref{eq:Uzrel}. Figures \ref{az240rR0.4mu1.0ref49_SEUzRel} and \ref{az270rR0.4mu0.85ref168UzRel} show the axial velocity surfaces and contours of the SEV. The increasing values of x-coordinate points in the direction of freestream, the increasing values of y-coordinate points towards the rotor blade, and the increasing negative $Uz_{rel}$ value points radially inwards on the rotor blade. The surface plot shows the prominent core axial velocities in comparison with the near far-field velocities. And the positive velocities in the near far-field shows the effect of centrifugal forces. The approximate size and location of the SEV are denoted by the dotted red circle on the contour.  

\begin{equation}\label{eq:Uzrel}
Uz_{rel} = w + U_{\infty}sin(\Psi)
\end{equation}

At azimuth 240 degrees, a strong axial velocity component is observed at inboard locations, thereby showing the growth of the helical vortex from the outboard location. The axial velocity increases with increase in advance ratio. The positive velocities outside the periphery of the SEV indicate the centrifugal forces in the near farfield of the rotor blade surface. Similarly, the azimuth 270 degrees and 300 degrees show a prominent axial velocity component, comparatively larger than those at the 240 degrees azimuth. With no spanwise component of the freestream at 270 degrees, the positive velocities outside the periphery of the SEV again show the presence of centrifugal stresses. This behavior is not significantly observed at 300 degrees azimuth.  

At azimuth 240 degrees, shown in Figure \ref{az240rR0.4mu1.0ref49_SEUzRel}, the core axial velocities increase radially inwards on the rotor blade. And a higher peak value is observed with the increase in advance ratio. In comparison to 240 degrees azimuth where there is a prominent peak in the core axial velocity, the broadening of this peak is observed at azimuth 270 degrees as shown in Figure \ref{az270rR0.4mu0.85ref168UzRel}. In contrast to 240 degrees, higher peak values are observed at advance ratios of 0.85, and there is no presence of a prominent peak at advance ratios of 1.0. At azimuth 300 degrees, a much broader peak was observed at the inboard radial locations. And for the radial stations greater than r/R 0.514 shown in Figure \ref{az300rR0.514mu0.7ref187.5UzRel}, the axial velocities at the surface were much higher than the core velocities. This can be observed from the position of the peaks on the surface with respect to the position of the dotted circle on the contour. 

\subsection{Life cycle of the SEV}

SPIV measurements were performed at smaller increments in azimuth around 240, 270, 290 and 300 degrees azimuth. The velocity fields were obtained at $\Psi \pm 5^{\circ}$ at several radial stations for different advance ratio. The average azimuthal gradient ($\frac{d\Gamma}{d\Psi}$) of the circulation is obtained from several radial locations at a given azimuth.  

\begin{figure}[!ht]
	\includegraphics[width=1\linewidth]{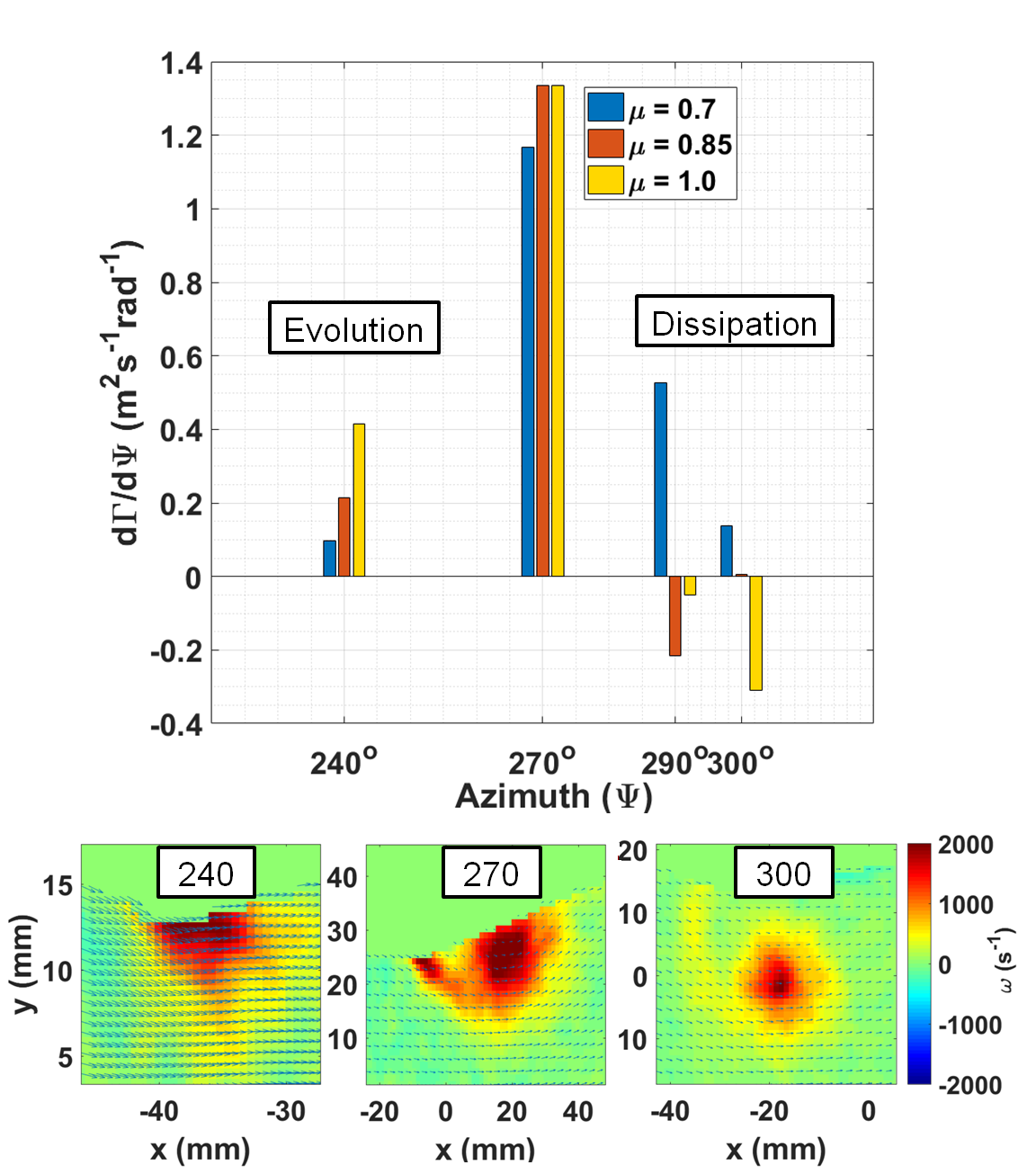} 
	\caption{Circulation gradient at various advance ratios}
	\label{CirculationAzimuth}
	
\end{figure}

Table \ref{Tab:CirculationGradient} and Figure \ref{CirculationAzimuth} show the azimuthal gradient of circulation. The reverse flow region before 270 degree azimuth can be considered as the ``evolution" phase and the region after 270 as the ``dissipation" phase. In the evolutionary phase, the gradient increases from 240 degrees to 270 degrees. The SEV evolves in this region due to the feeding of vorticity from the sharp edge. The higher advance ratios show higher gradients of circulation. This may be attributed to higher feed rates of vorticity from the sharp edge or to the transfer of kinetic energy from the mean flow into rotational energy of the vortex. Beyond 270 degrees azimuth is the dissipation phase wherein the SEV bursts or decays. The feed of vorticity from the sharp edge is not sufficient to sustain the vortex. It is counterbalanced by the factors assisting its decay. The counteracting factors are unknown at this point from this research. The circulation gradient in the dissipation region gives a metric for the earlier defined qualitative term ``history effects" attributed to the behavior of SEV beyond 270 degrees azimuth. The circulation gradients increase monotonically in the evolutionary phase with the increase in advance ratio, but such a monotonic behavior is not observed in the dissipation phase.      

\begin{table}[!ht]
	\centering
	\caption{Circulation gradient at various advance ratios}
	\begin{tabular}{ |c||c|c|c|  }
		\hline
       \multicolumn{4}{|c|}{$\frac{d\Gamma}{d\Psi}$ ($m^2s^{-1}rad^{-1}$)} \\
		\hline
		$\Psi$ & $\mu=0.7$ & $\mu=0.85$ & $\mu=1.0$ \\
		\hline
		$240^{\circ}$  & 0.097 $\pm$ 0.017 & 0.214 $\pm$ 0.052 & 0.416 $\pm$ 0.193  \\
		$270^{\circ}$  & 1.167 $\pm$ 0.161 & 1.335 $\pm$ 0.262 & 0.948 $\pm$ 0.424 \\
		$290^{\circ}$  & 0.527 $\pm$ 0.067 & -0.215 $\pm$ 0.806 & -0.051 $\pm$ 0.072 \\
		$300^{\circ}$  & 0.137 $\pm$ 0.114 & -0.005 $\pm$ 0.909 & -0.310 $\pm$ 0.385  \\
		\hline
	\end{tabular}
	
	\label{Tab:CirculationGradient}
\end{table}

\section{Conclusions}
The morphology of the sharp edge vortex observed in the reverse flow regime is established in this paper. The similarities with the leading edge vortex on a delta wing draws in attention towards the geometric sweep angle as a key feature to study the life cycle of a sharp edge vortex. The source of vortex generation could be due to the pitching and flapping rate of the rotor blade, but the sustenance of the vortex is dictated by the sweep effects and the differential on-set velocity. 

\begin{enumerate}
    \item The sharp edge vortex begins to form at the start of revere flow regime and grows radially inwards.
    \item The reverse flow envelope increases at higher advance ratios, and the attached vortex evolves as the rotor blade traverses in azimuthal direction. 
    \item The convection speed of the SEV shows parabolic behavior along the radius of the rotor blade. The convection speed ranges between $0.07U_{\infty} - 0.27U_{\infty}$.
    \item The vortex core size increases radially inwards. At the azimuth angles of 270 degrees and 300 degrees, the vortex core stretches all the way to the periphery of the vortex, showing the signs of a burst vortex.  
    \item The axial velocities in the core have a prominent peak in the evolving phase of the vortex. At azimuth angles of 270 degrees and 300 degrees, the axial velocity peak broadens. And the peak axial velocities shift from the vortex core towards the surface.
    \item The evolutionary phase shows monotonic increase in circulation gradient with the increase in advance ratio, and such a behavior is not observed in the dissipation phase. 
\end{enumerate}

\begin{acknowledgements}
	This work was funded by the US Army Research office under grant number W911NF1010398. Dr. Bryan Glaz Ferguson was the Technical Monitor. The technical POCs were Gloria Yamauchi and Oliver Wong. The authors are grateful to Jackson Merkl, Zhujia Huang in particular and the rest of the students in the Experimental Aerodynamics and Concepts Group for their assistance.

\end{acknowledgements}


\FloatBarrier
\bibliography{Bib.bib}
\bibliographystyle{plain}
\end{document}